\documentclass[a4paper,twocolumn,numberedappendix,twocolappendix,revtex4,iop]{openjournal}

\usepackage[T1]{fontenc}
\usepackage[utf8]{inputenc}
\usepackage[a4paper,marginpar=10pt,headheight=20pt, right=0.5cm, top=1.5in, left=2cm, bottom=0pt, footskip=0pt, footnotesep=10pt]{geometry}

\usepackage{booktabs}
\usepackage{bm}		
\usepackage[usenames,dvipsnames]{xcolor}
\usepackage[colorlinks,linkcolor=blue,citecolor=blue,urlcolor=blue ]{hyperref}
\usepackage{orcidlink}
\customauthor[Mittal, Oayda \& Lewis]{}
\customtitle[Multipoles-II]{}

\definecolor{valecol}{rgb}{0,0.5, 1.}

\def\apjl{ApJ Let.}

\def\aap{A\&A}

\def\apjs{ApJS}

\providecommand{\adsurl}[1]{\href{#1}{ADS}}
\providecommand{\eprint}[1]{\href{http://arxiv.org/abs/#1}{#1}}
\usepackage{ifthen}\def\eprinttmp@#1arXiv:#2 [#3]#4@{\ifthenelse{\equal{#3}{x}}{\href{http://arxiv.org/abs/#1}{#1}}{\href{http://arxiv.org/abs/#2}{arXiv:#2} [#3]}}
\renewcommand{\eprint}[1]{\eprinttmp@#1arXiv: [x]@}

\definecolor{maroon}{rgb}{0.5, 0, 0}
\definecolor{darkcyan}{rgb}{0.0, 0.55, 0.55}
\definecolor{overdensitycolor}{rgb}{1.0, 0.75, 0.0}
\definecolor{underdensitycolor}{rgb}{0.525, 0.424, 0.635}
\newcommand{\dotdeg}{\rlap{.}^\circ}
\usepackage{amssymb, graphicx, MnSymbol, multirow, subfig, ulem}

\begin{document}

\title{Cosmic Multipoles in Galaxy Surveys -- II.\\ 
        Comparing different methods in assessing the cosmic dipole}
\author{
    Vasudev Mittal,$^{1\star}\orcidlink{0000-0002-1708-6088}$
    Oliver T. Oayda,$^{1}\orcidlink{0009-0002-5013-8959}$
    and
    Geraint F. Lewis$^{1}\orcidlink{0000-0003-3081-9319}$
    }
\thanks{$^\star$ \href{mailto:vasudev.mittal@sydney.edu.au}{vasudev.mittal@sydney.edu.au}}
\affiliation{
    $^{1}$Sydney Institute for Astronomy, School of Physics A28, The University of Sydney, NSW 2006, Australia
    }

\begin{abstract}
We present a comparative analysis of estimators and Bayesian methods
for determining the number count dipole from cosmological surveys.
The increase in discordance between the number count dipole and the CMB's kinematic dipole has presented a challenge to the assumption of an isotropic and homogeneous universe.
The level of discordance has depended on the choice of method to determine the dipole; hence, there is a need to compare them to determine the better approach.
We select the \textsc{healpy} \texttt{fit\_dipole} algorithm as our estimator and show that it gives unbiased results regardless of the noise levels and sky coverage in the data.
However, for low sky coverage, the estimator's results have a large variance, which indicates that the estimator is not reliable in that regime.
We then compare the estimator's outputs with the Bayesian results.
If the sky coverage is sufficient, the Bayesian posterior probabilities agree with both the estimator's outcomes and the true parameter values.
But if the sky coverage is low, Bayesian analysis is often inconclusive, which safeguards against incorrect conclusions.
Both methods provide the ability to analyse multiple samples: on one hand, we need to select a different estimator, while in Bayesian inference, we need to extend our likelihood function to incorporate additional parameters.
Our study emphasises the need to perform a principled statistical analysis of sky surveys for dipole determination.
\end{abstract}

\maketitle

\section{Introduction}
The current cosmological framework---$\Lambda$CDM---rests on the assumption that the observed universe is homogeneous and isotropic at the largest scales.
Einstein first proposed this assumption of a spatially symmetric universe, which is currently known as the Cosmological Principle \citep[CP;][]{1917SPAW.......142E, milne1935}.
The most substantial support for the CP comes from the smoothness of the Cosmic Microwave Background (CMB), which has an average temperature of $2.7\,\text{K}$ \citep{planck2020}.
Small-scale fluctuations in the CMB are perceived to be the seeds of the current universe's large-scale structures \citep{Weinberg:2008zzc}.
However, these $10 \,\mu\text{K}$ scale fluctuations are dominated by a mK-scale dipole anisotropy over the sky.
The dipole anisotropy is interpreted as arising due to the motion of our heliocentric reference frame with respect to the `cosmic rest frame', where this anisotropy is absent.
This kinematic interpretation assumes the validity of the CP \emph{a priori}, with no underlying theoretical argument that explains the presence of a mK scale dipole anisotropy in an isotropic CMB \citep{aluri2023}.
If this view is accurate, then special relativity dictates that our motion should change the distribution of observables around us \citep{ellis1984}. 
Indeed, the all-sky distribution of sources such as radio galaxies \citep{blake2002} and quasars \citep{secrest2021} shows a dipole-like variation throughout the sky, which has been dubbed the matter dipole.
However, there is building evidence that the matter and kinematic dipoles are in disagreement with each other.
This `dipole tension' challenges the assumption of isotropy and is a pressing problem in modern cosmology \citep{Peebles_2022, Secrest:2025wyu, Secrest2025}.

A crucial step in measuring the matter dipole involves selecting an appropriate mathematical tool.
Historically, estimators have dominated the landscape, and many studies have devoted considerable efforts to selecting those that yield accurate and reliable results.
This is not a simple task, and inaccurate results arise from failing to take into consideration all potential sources of errors \citep{rubart2013, Secrest:2025wyu}.
This uncertainty has increased interest in Bayesian inference, where the focus shifts to testing a range of hypotheses by constructing appropriate likelihoods and priors.

In the analysis of the cosmic dipole, the choice of mathematical tool has been somewhat arbitrary, and a comparison between estimators and Bayesian inference has hitherto not been performed.
This study aims to fill this gap by comparing the results generated by both approaches.
Such a comparison requires the selection of an estimator whose reliability can be assessed by quantifying its bias and variance.
One such estimator is the \textsc{healpy}'s \texttt{fit\_dipole}, which implements a minimisation routine to calculate the dipole from a dataset.
While many recent studies \citep[such as][]{secrest2021, Kothari_2024} have tested it to be unbiased for their purposes, \texttt{fit\_dipole} has not been exhaustively tested on all monopoles, dipole directions/amplitudes, and sky masks.
Additionally, \citet{darling2022} performs a bias correction to their results, which throws its unbiasedness into question, necessitating a deeper enquiry.
Once we have answered this question and have understood the estimator's bias and variance, we can compare the statistical frameworks and determine which is best suited for dipole analysis.

The paper is set out as follows.
The next section briefly discusses the relevant literature.
Section \ref{sec: framework} discusses the mathematical framework behind the \textsc{healpy} estimator and Bayesian statistics.
We supplement our discussion on the estimator with Appendix \ref{sec:appendix}.
In Section \ref{sec: methodology}, we use simulations to study the statistical methods and deduce their important properties.
This section is accompanied by Appendix \ref{sec:estimator-bias-comparison}.
Section \ref{sec: discussion} discusses their implication on the choice of statistics.
Section \ref{sec: conclusion} concludes our findings.

\section{Background}\label{sec: background}
Under the kinematic interpretation, the mK scale dipole anisotropy of the CMB indicates that our solar system is moving towards the Galactic coordinates $(264\dotdeg021, 48\dotdeg253)$ with a speed of $369.82 \pm 0.11 \,\text{km}\,\text{s}^{-1}$ \citep{planck2020}.
The modulation of the number density of extra-Galactic sources due to this motion is calculated using a set of assumptions about the source population and special relativity \citep{ellis1984}:
\begin{itemize}
    \item The proportion of sources at low redshifts where the dipole is affected by clustering is insignificant.
    \item Within the observation passband, the source spectral energy distribution has a power law dependence on observed frequency, i.e. $S \propto \nu^{-\alpha}$.
    \item The cumulative number density of sources near the flux density limit has a power law dependence on the flux density ($N(>S) \propto S^{-x}$).
\end{itemize}
Then, up to the leading order, the observed number density contrast (in the direction $\bf{n}$) due to our motion at velocity $\bf{v}$ is 
\begin{equation}\label{eq:ellis}
    \frac{\Delta N ({\bf n})}{N} = [2+x(1+\alpha)] \frac{\bf v}{c}\cdot{\bf n}.
\end{equation}
This dipolar modulation was first observed in the NRAO VLA Sky Survey \citep[NVSS;][]{nvss-survey} where consistency with the CMB dipole, albeit with significant uncertainties, was reported \citep{blake2002}.\footnote{
    There are some previous studies such as \citet{Baleisis_1998}, where the kinematic dipole has been detected, but they used catalogues having $<\mathcal{O}(10^5)$ sources.
    NVSS was the first survey which generated a radio galaxy sample containing $\sim\mathcal{O}(10^5)$ sources.}
This study relied on a spherical harmonic decomposition of the observed number densities to calculate the dipole, and they used a $\chi^2$ analysis to compare the results with the CMB coefficients. 
This has been followed by several studies, where different levels of (dis)agreement between the matter and CMB dipole have been reported across different galaxy surveys \citep[some studies include][]{Crawford_2009, singal2011, rubart2013,colin2017, bengaly2018,secrest2021, secrest2022, darling2022}.
Each of these studies have used different estimators to measure the matter dipole.

Estimators are mathematical functions that take data as an input and return a point estimate of the desired property of the data (in our example, the properties of the dipole) as the output.
Imperfections in the data, such as incomplete sky coverage, noise, catalogue resolution, etc., can influence the estimator's outputs.
These imperfections impact the inferred dipole in two distinct ways: first, they bias the properties of the calculated dipole away from the true values.
Second, they increase the unpredictability of the estimator by increasing its variance.
Historically, many estimators have been used for dipole measures and often had to be corrected for their bias \citep[see the discussion in][]{singal2011, rubart2013, colin2017, siewert2021}.
Bias analysis is a complicated business; often, not all sources of bias are known beforehand, increasing the chances of obtaining incorrect results.
It is not uncommon to find that an estimator which was initially thought to be free of bias was actually biased.
A notable example is the linear estimator, which was initially used in \citet{Crawford_2009} without bias correction, but \citet{rubart2013} showed that it exhibited a complicated bias.
This obfuscates the analysis, and many studies have dedicated a significant portion of their text to bias quantification.

Some recent studies, however, have shifted towards using Bayesian inference for calculating the matter dipole in different galaxy surveys \citep{dam2023, wagenveld2023, mittal2024, oayda2024}.
Bayesian inference involves constructing the likelihood and prior functions for the hypothesis being studied, which are then used in Bayes's theorem to calculate the posterior probability distribution of the free parameters in the hypothesis or model.
Apart from determining the free parameters, Bayesian analysis also differs in its philosophy with regard to testing a particular model.
With most estimators, we calculate the probability of rejecting a hypothesis by first generating a large number of mock datasets using the hypothesis, and then determining the proportion of the mock datasets that give the results in agreement with the hypothesis.
In Bayesian statistics, instead we compute the odds ratio
between two (or potentially more) different models,
indicating which model is favoured over the other.

Given that no study has previously considered the choice of mathematical tool, there is a need to perform a general comparison between both methods to determine which is best-suited for cosmic dipole analysis.
We perform this comparison in this study and deduce the feasibility of either approach for calculating the matter dipole in galaxy surveys.

\section{Mathematical Framework}\label{sec: framework}
\subsection{$\textsc{Healpy}$ Estimator}
In this work, we focus on \textsc{healpy}'s \texttt{fit\_dipole} as the estimator.
This estimator calculates the dipole by implementing the least-squares minimization, which finds the peak of the likelihood function constructed by assuming that the observed data is sampled from an underlying Gaussian distribution.
The results generated using this method coincide with the peak of the Bayesian posterior distributions calculated using a Gaussian likelihood function and uninformative priors.
Even though this gives a Bayesian aspect to the estimator, in this work, we study it from a frequentist viewpoint, where we minimize the sum of squared residuals to generate a point estimate of the underlying dipole.

We choose to work with this estimator as there is no uniform consensus on the nature and extent of its bias, with different studies reporting varying outcomes.
Some studies have tested that this estimator gives unbiased results
\citep{secrest2021, Kothari_2024}, while others have corrected their results for a bias \citep{darling2022}.

As shown in some studies \citep{abghari2024, Kothari_2024}, this estimator calculates the dipolar modulation on a pixelated sphere through spherical harmonics, so its utility extends beyond cosmic dipole analysis. In particular, it implements a four-dimensional minimisation and calculates the free parameters $A_j$ in the following expression.
\begin{equation}\label{eq:estimator}
    U = \sum_{i}[\mathcal{N}_i - (A_{0}+A_{1}x_{i}+A_{2}y_{i}+A_{3}z_{i})]^{2}
\end{equation}
where $\mathcal{N}_i$ is the number density of the $i^{th}$ pixel centred at $(x_i,y_i,z_i)$ on the unit sphere, and the sum is taken over the unmasked pixels, i.e. pixels present in the observed part of the sky.
Here $A_0$ is the data's mean number density, and $(A_1, A_2, A_3)$ is the dipolar modulation of the mean number density.
The dipole vector modulating the catalogue is given by $\left( \frac{A_1}{A_0},\frac{A_2}{A_0},\frac{A_3}{A_0} \right)$.
Extremising $U$ with respect to $ A_j$ gives the following set of equations $(\mu_j = \{1,x, y, z\} ) $:
\begin{equation}
     \sum_{i}[\mathcal{N}_i \mu_{i} - (A_{0}\mu_{j}+A_{1}\mu_{j}x_{i}+A_{2}\mu_{j}y_{i}+A_{3}\mu_{j}z_{i})] = 0.
     \label{estimator-eqn-recasted}
\end{equation}
To constrain its properties, we need to discuss the estimator's behaviour for different number densities and sky coverages.
In discretised all-sky catalogues, a byproduct of binning a finite number of sources into different pixels is shot noise.
Since shot noise scales as $1/\sqrt{\text{Number Density}}$, a catalogue with a high number density will have proportionally less noise (with respect to the overall counts) as compared to the one with a lower number density.
It is crucial to study the estimator's behaviour in both situations.
In this study, we investigate two catalogue templates: one with minor/negligible noise (noiseless), and one with significant noise. 
As we show in \ref{sec:appendix}, the estimator gives the correct dipole for a noiseless sky regardless of sky coverage i.e. it is unbiased, and has negligible variance.
However, in the presence of noise, we find that the estimator's output does not agree with the true values.
Since real sky surveys always have noise in their number densities, it becomes important to understand the bias and variance structure for noise-injected catalogues. 
In this study, we will use simulations to understand these properties.
For this, we first need to understand how to calculate the bias and varince using simulations.

\subsection{Bias Characterization}
If we have an estimator $\hat{\phi}$ calculating the parameter $\phi$ from data, then the bias of the estimator is defined as 
\begin{equation}\label{eq:bias}
    \text{Bias}(\hat{\phi}) = \left<\hat{\phi}\right> - \phi_{\text{true}}
\end{equation}
where $\left<\space\right>$ denotes the expectation value of the estimator's outputs, calculated by taking the average over the dipoles determined from multiple noise-injected copies of a dataset (with $\phi_{\text{true}}$ as the true parameter).%
\footnote{It is important to note that the bias of an estimator is not a single value; rather, it depends on the properties of the dataset such as noise levels, shape and size of the mask, etc.}
In other words, bias is the shift of the estimator's expected outcome from the true value \citep{rubart2013}.
Although bias is a good indicator of an estimator's reliability, we must remember that calculating the expectation value might remove many vital features of the underlying distribution.
Hence, we will also study the entire distribution of the estimator's outcomes.

The \textsc{healpy} estimator gives the dipole vector $\left(\mathcal{D}_x, \mathcal{D}_y, \mathcal{D}_z \right)$ as the output, which we then convert into the dipole parameters $\left(\mathcal{D}, l^{\circ}, b^{\circ} \right)$ to facilitate comparison with the amplitude and direction of the expected kinematic dipole.
Here, the dipole amplitude $\mathcal{D}$ is calculated by taking the 2-norm of the dipole vector i.e. $\mathcal{D} = \sqrt{\mathcal{D}_x^2+ \mathcal{D}_y^2 + \mathcal{D}_z^2}$
Given this, there are two methods of performing the bias analysis:
\begin{enumerate}
    \item The first method is to average over the output dipole vectors $\left<\left(\mathcal{D}_x, \mathcal{D}_y, \mathcal{D}_z \right)\right>$, convert the average vector into dipole parameters and then compare it with true parameter values.
    Alternatively, we can compare $\left<\left(\mathcal{D}_x, \mathcal{D}_y, \mathcal{D}_z \right)\right>$ with $\left(\mathcal{D}_x, \mathcal{D}_y, \mathcal{D}_z \right)_{\text{true}}$ itself. 
    This method of studying the distribution of vector components has been used in \citet{Kothari_2024} with an extended version of the \textsc{healpy} estimator.
    \item The second method is to first convert the output dipole vectors to $\left(\mathcal{D}, l^{\circ}, b^{\circ} \right)$, take an average over $\mathcal{D}$ and $(l^{\circ}, b^{\circ})$ separately and then compare these averages with $\left(\mathcal{D}, l^{\circ}, b^{\circ} \right)_{\text{true}}$.
    This method of studying the distributions of amplitudes and directions separately has been used in studies such as \citet{secrest2021, siewert2021} for different estimators like the \textsc{healpy} and $\chi^2$ estimators.
    We use \citet{siewert2021}'s method for calculating the average over $(l^{\circ}, b^{\circ})$.
    First, we convert the calculated dipole vectors to unit vectors by dividing them by their amplitudes, then we calculate the average of these unit vectors, and finally, we convert this averaged vector to a mean direction $\left< (l^{\circ}, b^{\circ}) \right>$
\end{enumerate}
 At face value, we might anticipate that both methods give the same results regarding expected outcomes and, hence, have identical bias. 
 But we should remember that averages generally are not propagative (the average of any variable $\alpha$ does not characterise the average of any function of $\alpha$).
Hence, we will look at both methods and compare them with each other and Bayesian inference.

\subsection{Variance Characterization}
The variance of an estimator $\hat{\phi}$ is defined as
\begin{equation}
    \text{Variance}(\hat{\phi}) = \left<\hat{\phi}  - \left<\hat{\phi}\right> \right> 
\end{equation}
Similar to bias, the expectation values are calculated by averaging over the dipoles determined from multiple copies of the noise-injected datasets. 
It measures the spread of outputs given by the estimator.
Additionally, it also quantifies the confidence that we have in the estimator's outputs.
One popular method of measuring the confidence in an estimator's results is to use Confidence Intervals (CI).
A CI is a range of values within which the true value of the estimated parameter can lie with a specified confidence.
CIs are proportional to $\sqrt{\text{Variance}}$, which means that an estimator with a lower variance is better suited for analysis.

Taken together, the bias and variance of an estimator are used to assess its reliability.
Ideally, an estimator should have both low bias and low variance to ensure accurate and stable inference. 
However, in practice, especially with limited data or high-dimensional problems, it may be preferable to accept a small, well-understood bias in exchange for reduced variance. 
The priority is to maintain control and interpretability of the estimator’s behaviour rather than minimising bias alone.

\subsection{Bayesian Inference}
Our Bayesian analysis is the same as in the companion study \citep{Oayda:2024voo} and other works \citep{mittal2024, oayda2024}. Bayesian analysis uses the Bayes theorem to calculate the probability distribution of a model's (M) parameters $\Theta$ given the dataset $\mathbf{D}$:
\begin{equation}
    P( \Theta \, | \, \mathbf{D}, M )
        = \frac{\mathcal{L}(\mathbf{D} \, | \, \Theta, M) \pi(\Theta \, | \, M)}
        {\mathcal{Z}(\mathbf{D} \, | \, M)}. \label{eq:bayes-theorem}
\end{equation}
We incorporate our beliefs about the data into the analysis through the priors probabilities $\pi$.
To define our likelihood function, we assign a scalar function $f(\mathbf{\hat{p}}_i)$ to the $i$-th sky pixel pointing in the direction $\mathbf{\hat{p}}_i$. 
The exact form of $f$ depends on the model that we want to investigate.
In order to convert this to probabilities, we normalize $f$ by dividing it by its sum over the whole sky i.e.
\begin{equation}
    \hat{f}(\mathbf{\hat{p}}_i) = \frac{%
                    f(\mathbf{\hat{p}}_i)}{\sum_{i=1}^{n_{\text{pix}}} f(\mathbf{\hat{p}}_i)}
\end{equation}
This normalized function denotes the probability of observing a single source in the $i$-th pixel.
Since each pixel has multiple sources, the net probability of observing $\mathcal{N}_i$ sources in the $i$-th pixel is proportional to $[\hat{f}_{i}]^{\mathcal{N}_i}$
Using this, we construct our likelihood function as follows:
\begin{equation}
    \ln \mathcal{L} = \sum_{i=1}^{n_{\text{pix}}} \mathcal{N}_i \ln \hat{f}(\mathbf{\hat{p}}_i) = 
        \sum_{i=1}^{n_{\text{pix}}} \mathcal{N}_i
        \ln \left( \frac{%
                    f(\mathbf{\hat{p}}_i)}{\sum_{j=1}^{n_{\text{pix}}} f(\mathbf{\hat{p}}_j)}
            \right).
        \label{eq:pbp-likelihood}
\end{equation}
This approach of studying each source individually was first used for determining distances using the tip of the red giant branch of a sparsely populated system \citep{conn2011,conn2012}
We use the following models for our analysis
\begin{enumerate}
    \item $M_0$: Monopole (null hypothesis), with $f = 1$.
    \item $M_1$: Combination of a monopole and a dipole, with $f = 1 + \mathcal{D} \cos \theta_i$.
    Here $\theta_i$ is the angular offset between the $i$-th sky pixel and the dipole direction.
\end{enumerate}
Our priors for the dipole parameters $(\mathcal{D},l,b)$ are sampled from the following distribution.
\begin{enumerate}
    \item $\mathcal{D} \sim \mathcal{U}(0,1)$.
    \item $l \sim \mathcal{U}(0, 2\pi)$.
    \item $b \sim \cos^{-1}(1 - 2u)$ for $u \sim \mathcal{U}(0,1)$.
\end{enumerate}
Here $\mathcal{U}$ denotes a uniform distribution within the mentioned limits.
This prior choice is motivated from the principle of indifference,
reflecting our lack of information about the underlying dipole in the data.

For hypothesis comparison, we use the odds ratio for two models $M_1$ and $M_2$
\begin{equation}
    \frac{P(M_1 \, | \, \mathbf{D})}{P(M_2 \, | \, \mathbf{D})}
           = \frac{\pi(M_1)}{\pi(M_2)}
             \frac{\mathcal{L}(\mathbf{D} \, | \, M_1)}
                {\mathcal{L}(\mathbf{D} \, | \, M_2)}
            = \pi_{12} B_{12}
\end{equation}
where $\pi_{12}$ represents our belief about the relative strength of the models and $B_{ij}$ is the Bayes factor, which gives information about the relative predictive powers of the models.
We use the natural logarithm of the Bayes factor
\begin{equation}
    \ln B_{i0} = \ln \mathcal{Z}_i - \ln \mathcal{Z}_0
\end{equation}
along with Jeffreys's scale \citep{kass1995} to interpret the values qualitatively.
We use \textsc{dynesty} \citep{dynesty-v2.1.3} \footnote{\url{https://pypi.org/project/dynesty/}}, a \textsc{python} package, for calculating the posterior distributions and the Bayes factors.
\textsc{dynesty} implements the Nested Sampling algorithm, where posteriors are sampled in shells of increasing likelihood \citep{skilling2004, skilling2006}.

\section{Simulations}\label{sec: methodology}
\subsection{Methodology}
We construct our mock catalogues using the method discussed in the companion study \citet{Oayda:2024voo}, which we will briefly mention again. 
We first pixelate the sky into equal area pixels using the \textsc{healpix}\footnote{\url{https://healpix.sourceforge.io/}} algorithm in \textsc{Python}, through the \textsc{healpy} package \citep{Gorski2005, Zonca2019}.
We use \texttt{nside=64} to generate $49152$ equi-area pixels.
Next, we assign a mean number density $\bar{\mathcal{N}}=50$ to each pixel, which gives us $\approx2.5$ million sources over the unmasked sky.
Then, we modulate the number densities using a dipole of amplitude $\mathcal{D}=0.007$ pointing in the direction of CMB dipole, i.e. $(l,b)=(264\dotdeg021,48\dotdeg253)$.
This modifies the number density of a pixel from $\bar{\mathcal{N}}$ (monopole) to $\bar{\mathcal{N}}(1+\mathcal{D} \cos \theta_i)$ (combination of monopole and dipole)
These template dipole parameters are identical to those used in our previous work \citet{Oayda:2024voo}, and are consistent with the expected kinematic dipole amplitude for quasars \citep{secrest2021}.

We refer to these catalogues as the template catalogues.
These template catalogues are then injected with shot noise through Poisson sampling to generate our mock catalogues.
More specifically, the final number count density $\mathcal{N}_i$ for each pixel is sampled from a Poisson distribution 
\begin{equation}
    P( \mathcal{N}_i \, | \, \lambda_i )
        = \frac{\lambda_i^{\mathcal{N}_i} e^{-\lambda_i}}{\mathcal{N}_i!}
\end{equation}
where the rate parameter is set to the density of the template catalogue, i.e. $\lambda_i = \bar{\mathcal{N}}(1+\mathcal{D} \cos \theta_i)$.

Since astronomical surveys rarely probe the complete sky uniformly, we mask our mock catalogues to mimic their sky coverage.
Currently, the quasar catalogues used in cosmic dipole studies are constructed from observations made in visible and infrared bands using space-based telescopes.
Consequently, their observations in regions around the Galactic plane are affected by dust.
On the other hand, since radio surveys are ground-based, they cover a single equatorial hemisphere.
Another possibility to consider is the situation where the survey observes a number of sky patches which might or might not overlap.
In our previous paper \citep{Oayda:2024voo}, we have shown that it is possible to determine the number-count dipole from such surveys using Bayesian inference, provided that the number density of the catalogue is sufficient.
Here, we will investigate whether we can use the estimator to determine the dipole in such a scenario.
Accordingly, we superimpose each mock catalogue with the following masks.
\begin{enumerate}
    \item \emph{Galactic plane masks:} We mask the sky between the latitudes $(-b^{\circ}, b^{\circ})$, where $b^{\circ}$ is selected from the range of $0^{\circ}$ to $80^{\circ}$  in increments $10^{\circ}$.
    This mask is symmetric about the galactic plane.
    We use $g_{\text{mask}}$ to denote the masking latitude $b^{\circ}$.
    \item  \emph{Equatorial polar cap:} We mask the regions around the southern equatorial pole using circular caps centred at the pole.
    The angular radius of the cap is selected from the range of $15^{\circ}$ to $150^{\circ}$ in increments of $15^{\circ}$.
    We use $\Delta_{s}$ to label the angular radius of the circular cap, with sub-script $s$ referring to a mask centred on the south equatorial pole
    The north-equatorial pole mask $\Delta_{n}$ is functionally equivalent to this mask and hence, we do not study that case separately.
    \item \emph{Discontinuous Survey:} We follow \citet{Oayda:2024voo}'s strategy of defining the sky-coverage of a discontinuous survey by selecting a fixed number of sky-patches below a given declination.
    We achieve this by first selecting $391$ sky pixels below declination $-10^\circ$.
    For each selected pixel, we include all the neighbouring pixels in our list of observed pixels
    (This means that each sky-patch has an area of $\sim 4.2\,\text{deg}^{2}$).
    Similar to \citet{Oayda:2024voo}, we increase the mean number density of the catalogue to $\bar{N}\approx2300$.
\end{enumerate}
To analyse the estimator, we generate a total of $100\,000$ mock catalogues.
We then superimpose each catalogue with one of the aforementioned masks and set the values of masked pixels to \texttt{UNSEEN}, a special \textsc{healpix} floating point value assigned to the masked pixels.
Next, we use the \texttt{fit\_dipole} function to calculate the dipole on each catalogue.
This procedure generates $100\,000$ data points for each type of mask.
Finally, we calculate the expectation values for the output dipole vectors and inferred dipole parameters to make deductions about the estimator's properties.
While performing Bayesian analysis, we explicitly remove all the masked pixels from the analysis and then calculate the posterior distribution using Nested Sampling.
We analyse $100$ mock catalogues for each mask and sample $1000$ data points from each posterior distribution.
We then aggregate all the sampled data points to obtain a single consolidated distribution for one specific mask type, which contains $ 100,000$ data points.
Ideally, for each mask type, presenting the result of a single nested sampling run is sufficient, yet we consolidate the results over $100$ iterations to show that the posterior distribution produced from the nested sampling runs is robust against noise.

\subsection{Results: Galactic plane masks}

\subsubsection{Estimator results}
\begin{figure*}
    \centering
    \subfloat[{\centering Distribution of output dipole amplitudes calculated \\ using the \textsc{Healpy} estimator}]{
    \includegraphics[width=0.4\linewidth]{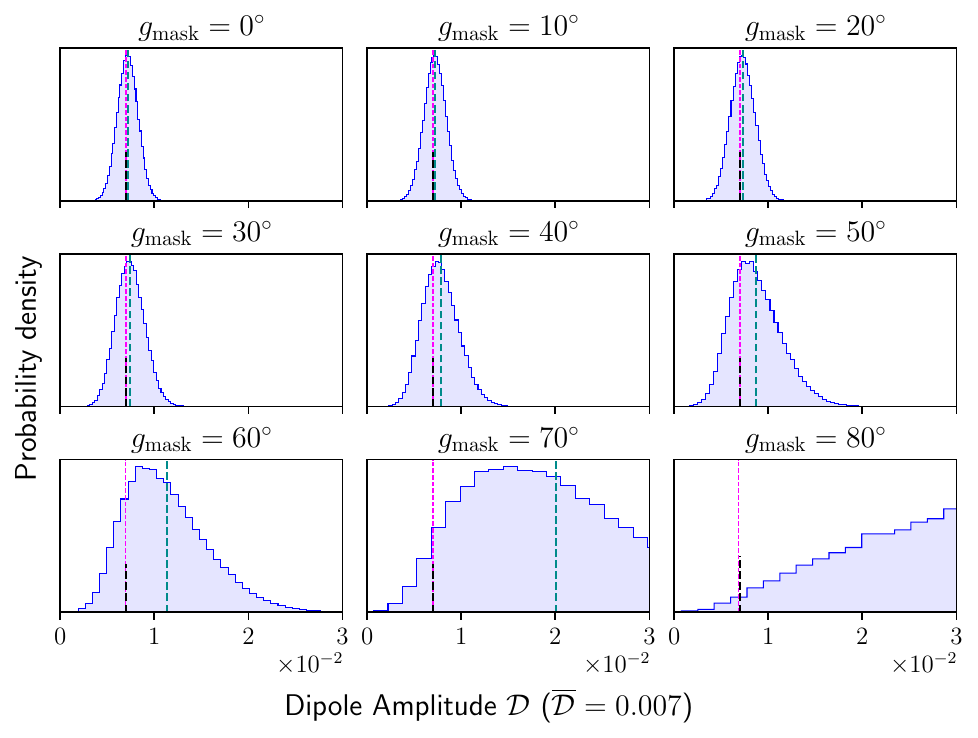}}\hfill
    \subfloat[{\centering Distribution of output dipole directions calculated \\ using the \textsc{Healpy} estimator}]{
    \includegraphics[width=0.58\linewidth]{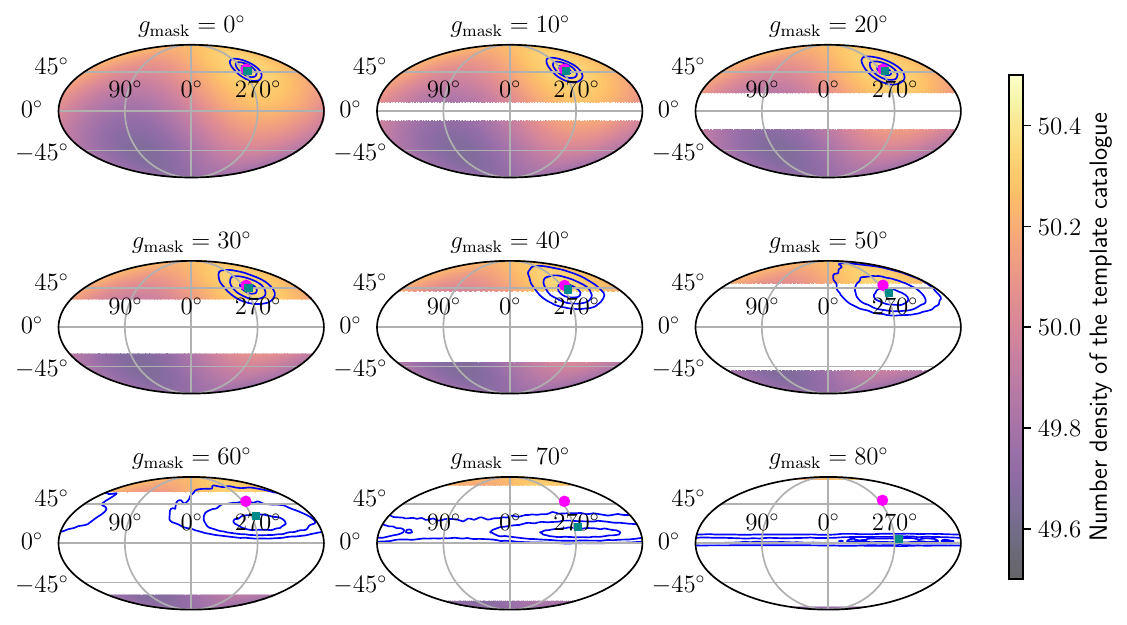}}
    \captionsetup{font=footnotesize}
    \caption{Evolution of the dipole amplitude and direction distributions inferred by the \textsc{healpy} estimator with increasing $g_{\text{mask}}$, shown in {\color{blue} blue}.
    Direction projection contours enclose $\approx 12\%$, $\approx 39\%$ and $\approx 69\%$ of the distribution.
    Expectation values for the estimator results are also shown.
    Expected dipole parameters calculated using the Method 1 are shown in {\color{magenta} magenta}, while the parameters calculated using the Method 2 are shown in {\color{darkcyan} dark cyan}.
    The black line marks the true dipole amplitude $0.007$..}
    \label{fig:galmask}
\end{figure*}
In Figure \ref{fig:galmask}, we show the distribution of dipole amplitudes and directions inferred from mock skies using the \textsc{Healpy} estimator.
We have overlaid the direction distributions on the masked sky template used to generate the mock catalogues. 
Note that the maximum number density contrast in the template catalogue is $\approx 1$ source per pixel.
When the sky coverage is high, we see that the amplitude and direction distributions are centred at the true values, and the variance in the distributions is quite low.
If we increase the $g_{mask}$ size (i.e. decrease the sky coverage), we see that the amplitude distribution first starts to develop a tail towards higher amplitude, and subsequently, the whole distribution shifts towards the high amplitude region.
Similarly, with a decrease in sky coverage, we see that the inferred direction distribution starts to flatten in a strip just above the galactic equator.
This broadening of the distributions also implies that the variance of the estimator increases significantly.
The expected dipole parameters for different averaging methods show the following trends:
\paragraph{Method 1:}
We show the dipole parameters calculated from the average dipole vector in Fig \ref{fig:galmask} using magenta markers.
The calculated amplitude agrees with the true dipole amplitude of $0.007$ for all Galactic masks, indicating no bias in amplitude.
Similarly, the calculated dipole direction aligns with the CMB dipole direction for all Galactic masks, indicating no bias in direction either.
\paragraph{Method 2:}
The average dipole parameters are marked in Fig \ref{fig:galmask} using dark cyan markers (these values correspond to the averages of the amplitude and direction distribution).
For smaller Galactic plane masks, the average amplitude is almost equal to the true dipole amplitude of $0.007$.
Meanwhile, the average direction aligns with the CMB dipole direction.
Increasing the mask size shifts the average amplitude towards higher values while the average direction shifts towards lower latitudes without much effect on their longitudes.
In general, an increase in the estimator's bias towards higher amplitude and lower latitudes is seen with an increase in the Galactic plane mask angle.

\subsubsection{Bayesian outcomes}
\begin{figure}
    \centering
    \includegraphics[width=\linewidth]{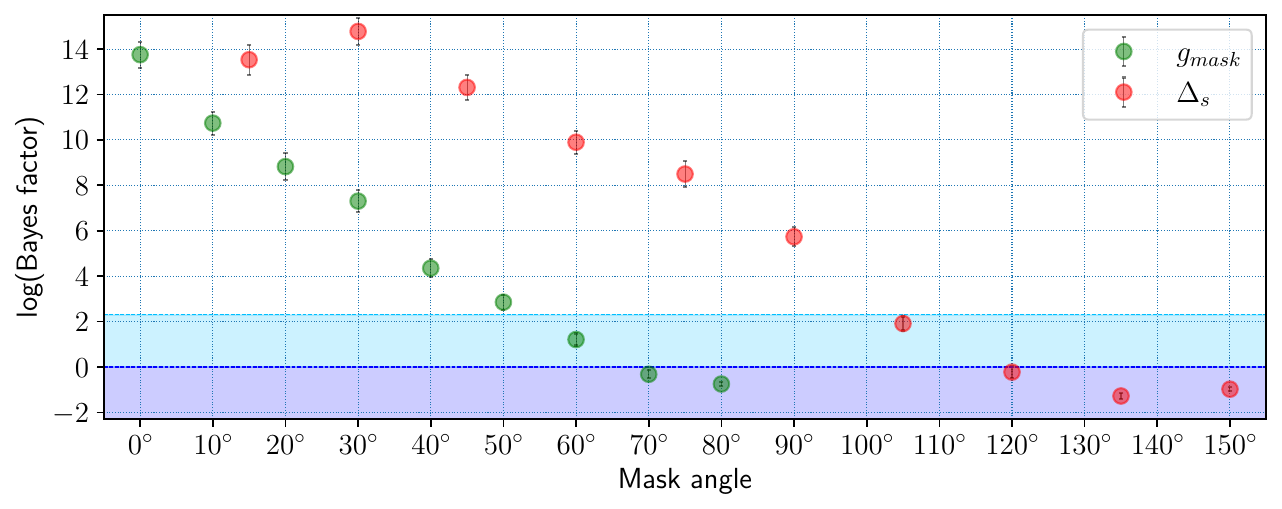}
    \captionsetup{font=footnotesize}
    \caption{Evolution of Bayes factors (averaged over 100 nested sampling runs) for dipole vs monopole with increasing mask size.
    \textcolor{Green}{Green} markers represent the average Bayes factors for the Galactic plane masks, while \textcolor{Red}{red} markers indicate the average Bayes factors for equatorial cap masks.
    Dashed lines bordering the shaded regions indicate the limits of no support ($\ln B_{10} = 0$) and strong support ($\ln B_{10} = 2.3$) for the presence of the dipole.
    The support values are as per Jefferys's scale.}
    \label{fig:galbayes}
\end{figure}
Fig \ref{fig:galbayes} shows the evolution of the logarithmic Bayes factors (for dipole vs monopole) with increasing $g_{\text{mask}}$ using green markers.
The posterior PDFs for the calculated dipole amplitude and direction distributions are shown in Figures 7--8 of \citet{Oayda:2024voo}.
With low Galactic plane masks, the peak of the recovered amplitude distribution coincides with the true dipole amplitude, and the direction distributions are concentrated around the CMB dipole direction.
Additionally, the Bayes factors show overwhelming support for the presence of the dipole. 
Increasing the size of the Galactic plane mask is equivalent to reducing the number of observed pixels and, by extension, the amount of information available for analysis. 
Accordingly, we see a decrease in the Bayes factors and a broadening of the posterior distributions.
Interestingly, the amplitude distributions develop a tail in the high-amplitude regions.
Meanwhile, the direction distributions become reasonably broad, with $39 \%$ and $69\%$ contours covering most of the northern sky.
The Bayes factors also become negative, implying that the analysis becomes indifferent to both the dipole and monopole hypotheses.

\subsection{Results: Equatorial cap masks}
\begin{figure*}[!ht]
    \centering
    \subfloat[{\centering Distribution of output dipole amplitudes calculated \\ using the \textsc{Healpy} estimator}]{
    \includegraphics[width=0.48\linewidth]{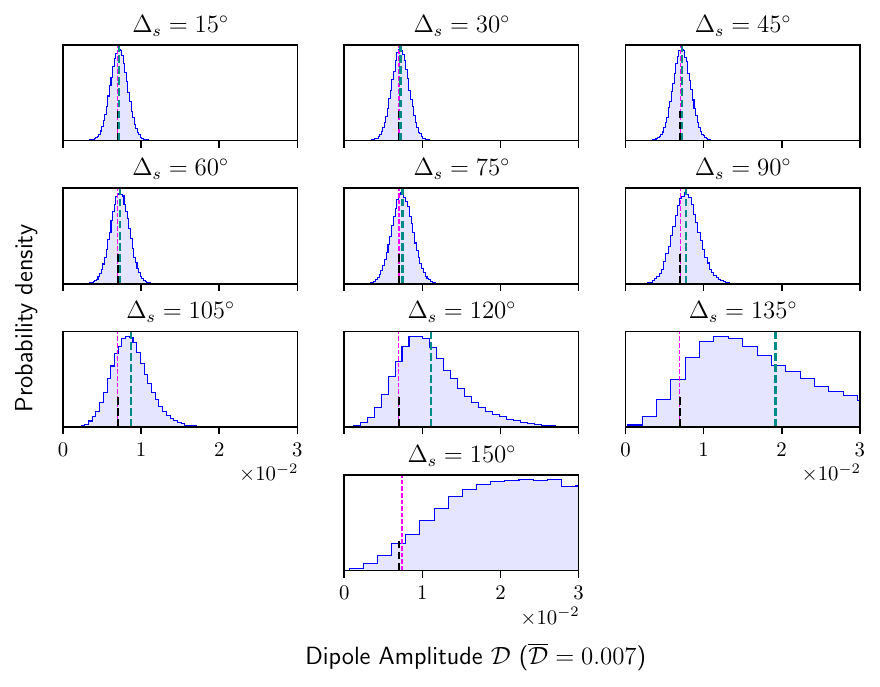}}\hfill
    \subfloat[{\centering Dipole amplitude PDFs calculated \\ using Bayesian inference}]{
    \includegraphics[width=0.48\linewidth]{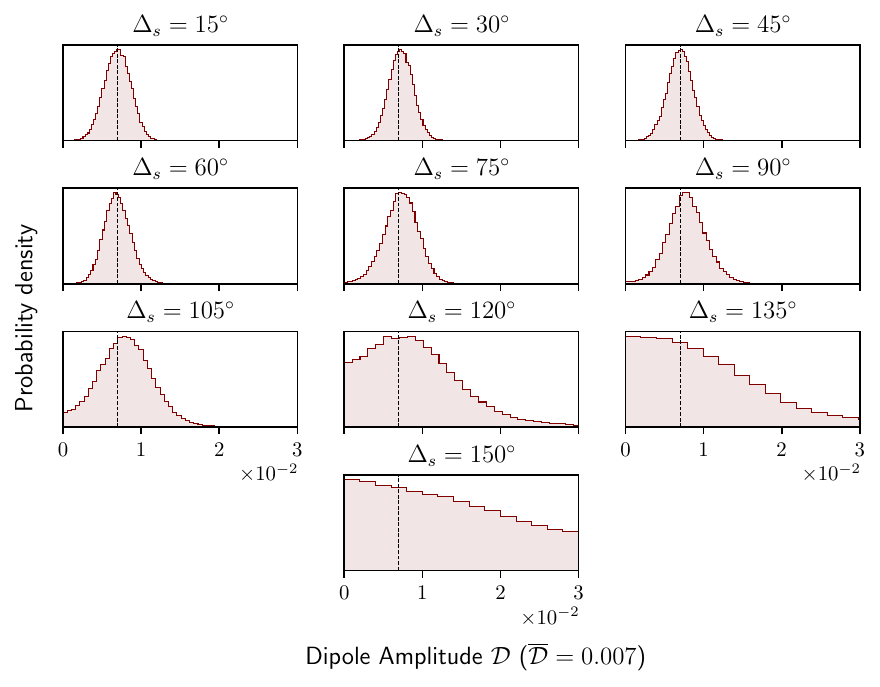}}\\
    \subfloat[{\centering Distribution of output dipole directions calculated \\ using the \textsc{Healpy} estimator}]{
    \includegraphics[width=0.49\linewidth]{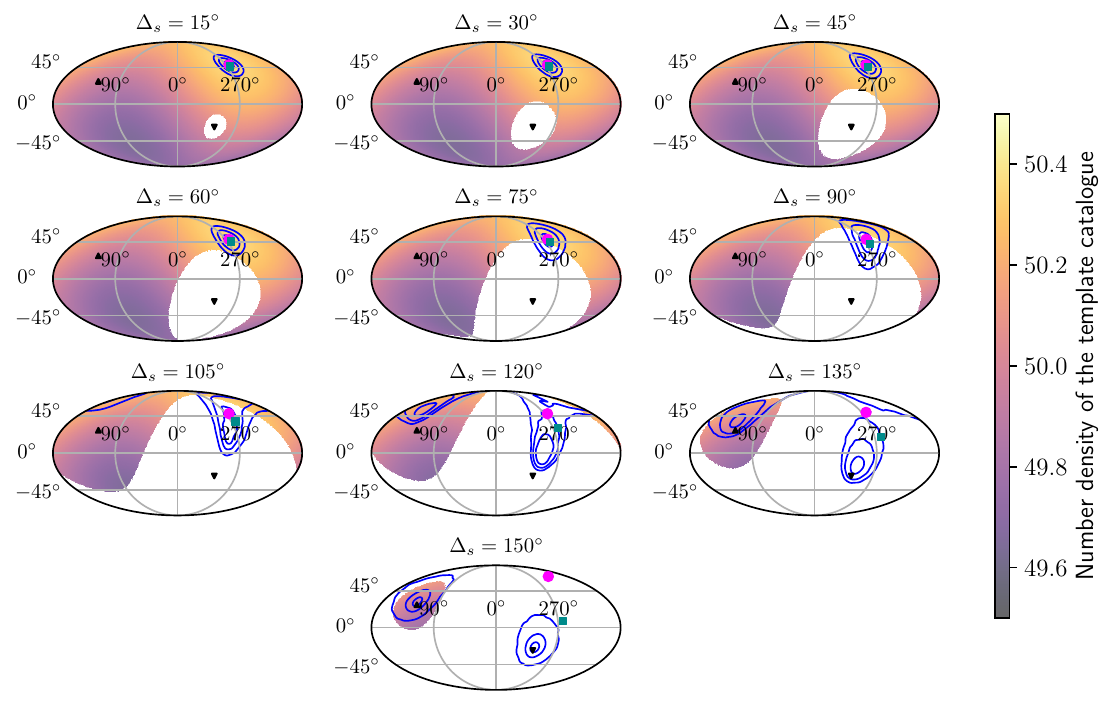}}\hfill
    \subfloat[{\centering Inferred dipole direction PDFs calculated \\ using Bayesian inference}]{
    \includegraphics[width=0.49\linewidth]{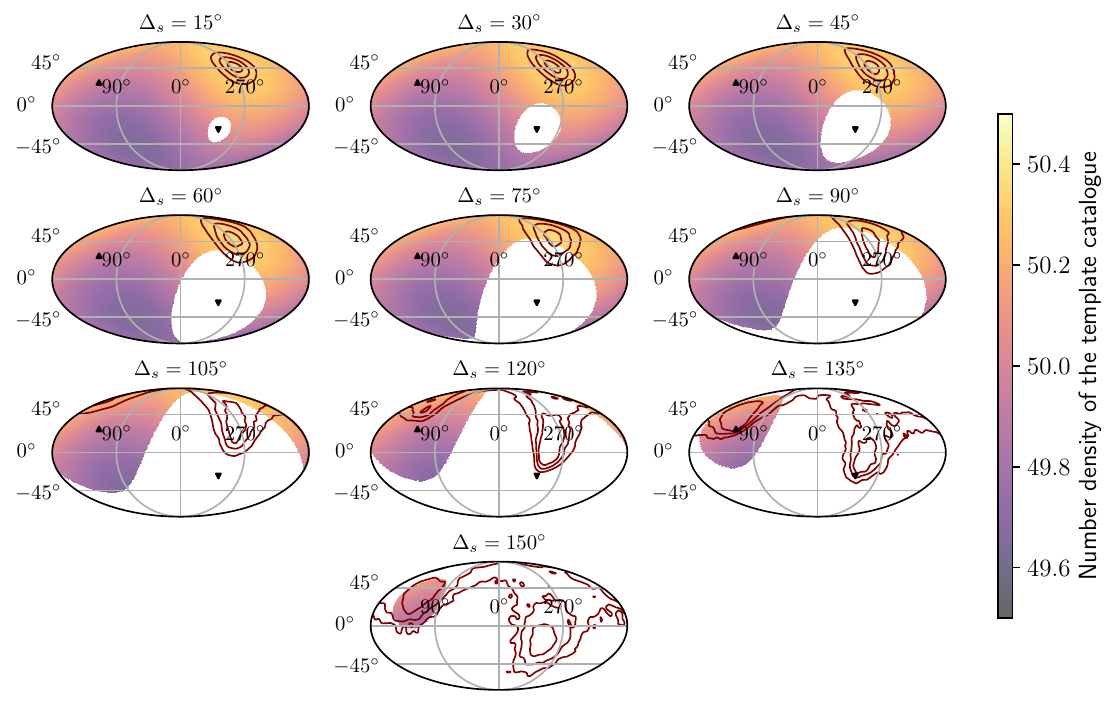}}
    \captionsetup{font=footnotesize}
    \caption{Evolution of the dipole amplitude and direction distributions inferred by the estimator (shown in {\color{blue} blue} ), and the amplitude and direction posteriors inferred by Bayesian analysis (shown in {\color{maroon} maroon} ) with increasing $\Delta_{s}$.
    Direction projection contours enclose $\approx 12\%$, $\approx 39\%$ and $\approx 69\%$ of the distribution for the estimator outputs and demarcate the corresponding credibility intervals for the Bayesian posteriors.
    Expectation values for estimator results are also shown.
    Refer to Fig \ref{fig:galmask} for interpretation of the visual elements.}
    \label{fig:eqmask}
\end{figure*}
\subsubsection{Estimator results}
In the left panels of Figure \ref{fig:eqmask}, we show the distribution of dipole amplitudes and directions inferred from mock skies using the estimator.
Similar to the results derived for galactic plane masks, when the sky coverage is high, we see that the amplitude and direction distributions are centred at the true values.
Again, the variance of the estimator is quite low in this regime.
Similarly, with an increase in $\Delta_{s}$, we see that the amplitude distribution first starts to develop a tail towards higher amplitude, and subsequently, the whole distribution shifts towards the high amplitude region.
The distribution of inferred directions, however, follows a different trend.
With an increase in $\Delta_{s}$, we see that the inferred direction distribution bifurcates into two separate distributions centred at the two equatorial poles, with the contours becoming disjointed, implying that its variance increases significantly.
The expected dipole parameters for different averaging methods show the following trends:
\paragraph{Method 1:}
Figure \ref{fig:eqmask} shows the dipole parameters calculated from average dipole vectors with magenta markers.
Similar to the results for Galactic plane masks, the dipole amplitudes and directions align with the true values, indicating the absence of bias in almost all scenarios.
Some directional offset is observed for the $150^{\circ}$ masks, with the calculated directions shifting slightly towards the unmasked patch of the sky.
\paragraph{Method 2:}
The average dipole parameters are shown in Figure \ref{fig:eqmask} using dark cyan markers.
For a low radius mask, like the Galactic plane masks, the average amplitude is almost equal to (but greater than) the true amplitude, along with a close alignment between the average and CMB dipole directions.
Increasing the mask size increases the bias in both amplitude and directional offsets.
The average amplitude becomes larger than the true value, and the average direction shifts to lower latitudes.
This trend is consistent for both types of equatorial masks.

\subsubsection{Bayesian outcomes}
The right panels of Figure \ref{fig:eqmask} shows the calculated posterior distributions for the dipole parameters while Figure \ref{fig:galbayes} shows the evolution of logarithmic Bayes factors with increasing $\Delta$.
We use red markers to denote their values.
Similar to the Galactic plane masks, the support for the presence of a dipole decreases (the Bayes factor decreases) with an increase in the mask size.
For masks with $\ln B_{10} \gtrsim 2.3$ (strong support for the presence of dipole), the peak of the amplitude posteriors coincide with the true value, and directions distributions are centred at the CMB dipole direction.
As the support for the dipole decreases, the amplitude distributions lose their sharp peak and show an increasing tail at high amplitudes.
At the same time, the direction posteriors evolve from a single distribution centred at the CMB dipole direction to a bimodal distribution, with their peaks pointing in the direction of both equatorial poles.
In addition, the $12\%$  and $39\%$ contours become disjointed but remain circumscribed by a single $69\%$ contour.

\subsection{Discontinuous survey}
\begin{figure*}[!ht]
    \centering
    \subfloat[Dipole Amplitude Distribution]{
    \includegraphics[width=0.25\linewidth]{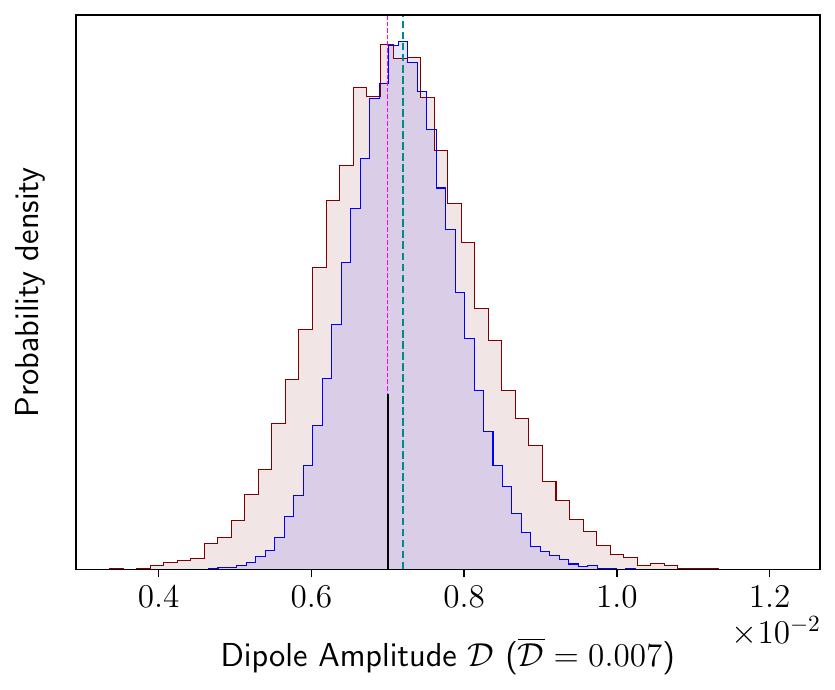}}\hfill
    \subfloat[Estimator's Inferred directions]{
    \includegraphics[width=0.33\linewidth]{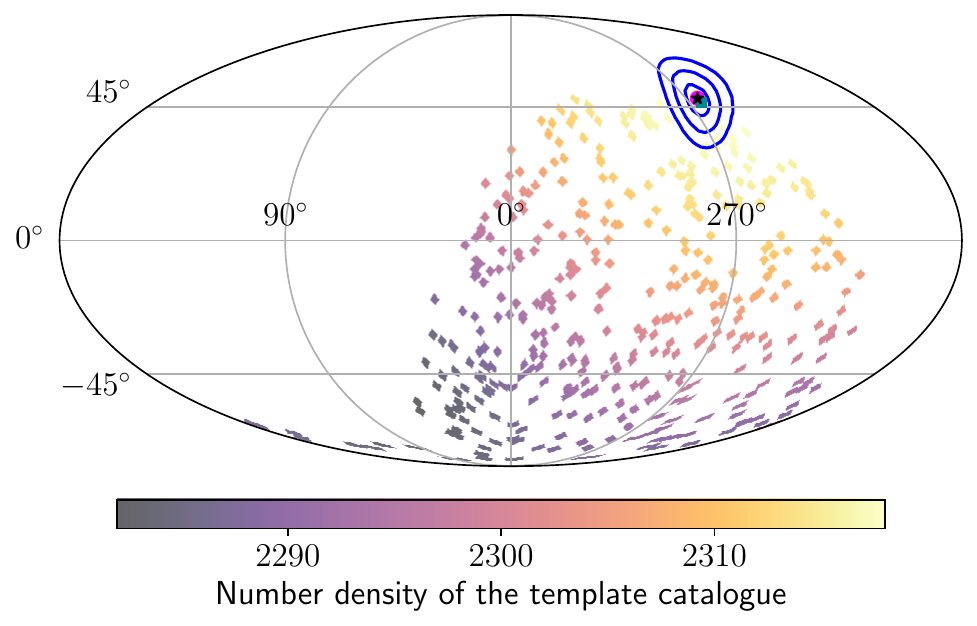}}\hfill
    \subfloat[Bayesian Direction Posteriors]{
    \includegraphics[width=0.33\linewidth]{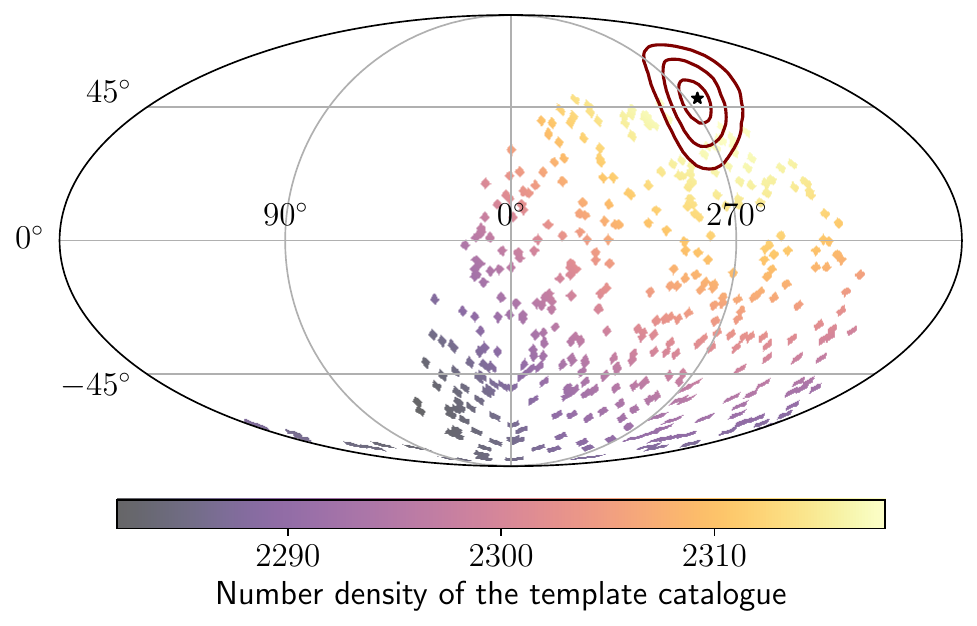}}\hfill
    \captionsetup{font=footnotesize}
    \caption{Dipole amplitude and direction distributions for the estimator outputs (shown in \textcolor{blue}{blue}), and posterior PDFs for the Bayesian outcomes (shown in \textcolor{maroon}{maroon}) for a discontinuous survey.
    Refer to Figs \ref{fig:galmask} and \ref{fig:eqmask} for interpretation of the visual elements.}
    \label{fig:discontinuousmask}
\end{figure*}
\subsubsection{Estimator results}
The blue histogram in the first plot in Figure \ref{fig:discontinuousmask} shows the distribution of inferred amplitudes calculated for the simulated sky.
Meanwhile, the second plot shows the distribution of inferred directions, which have been overlaid on the sky coverage map.
We see that both the amplitudes and direction distributions have low variance, and are centred at the true values.
Similarly, the expected parameters calculated using both the averaging methods (shown using fuchsia and dark-cyan markers) are in good agreement with the true parameter values.
This indicates that for a sufficiently high number density, the estimator's outputs are unbiased regardless of the choice of sky coverage or averaging method.

\subsubsection{Bayesian outcomes}
We find overwhelming evidence for the presence of a dipole in the simulated skies, with a logarithmic Bayes factor of $\approx 42$ in favor of the dipole model over the monopole.
The amplitude and direction posteriors are shown in the first and third plot using maroon histograms.
We find that both the posterior distributions are well-constrained and centred at the true parameter values.
This agrees with the results seen by us in our previous study \citep{Oayda:2024voo}.

\section{Discussion}\label{sec: discussion}
Based on the results presented above (and discussed in detail in \ref{sec:estimator-bias-comparison}), we find that the \textsc{healpy} estimator is unbiased in its native parametrisation (Cartesian $\mathcal{D}_x$, $\mathcal{D}_y$ and $\mathcal{D}_z$).
That is to say, computing the expectation of the dipole vector $\left< \mathcal{\mathbf{D}} \right>$, which amounts to summing the Cartesian vector outputs from many mock skies, yields a mean vector consistent with the sample truths.
It might be argued that the bias encountered in the dipole parameters is entirely due to our method for calculating the bias.
However, the parameter space (spherical $\mathcal{D}$, $l^{\circ}$ and $b^{\circ}$) distribution itself is not a good choice for studying the estimator's properties.

\begin{figure*}
    \centering
    \subfloat[Dipole vectors]{
    \includegraphics[width=0.49\linewidth]{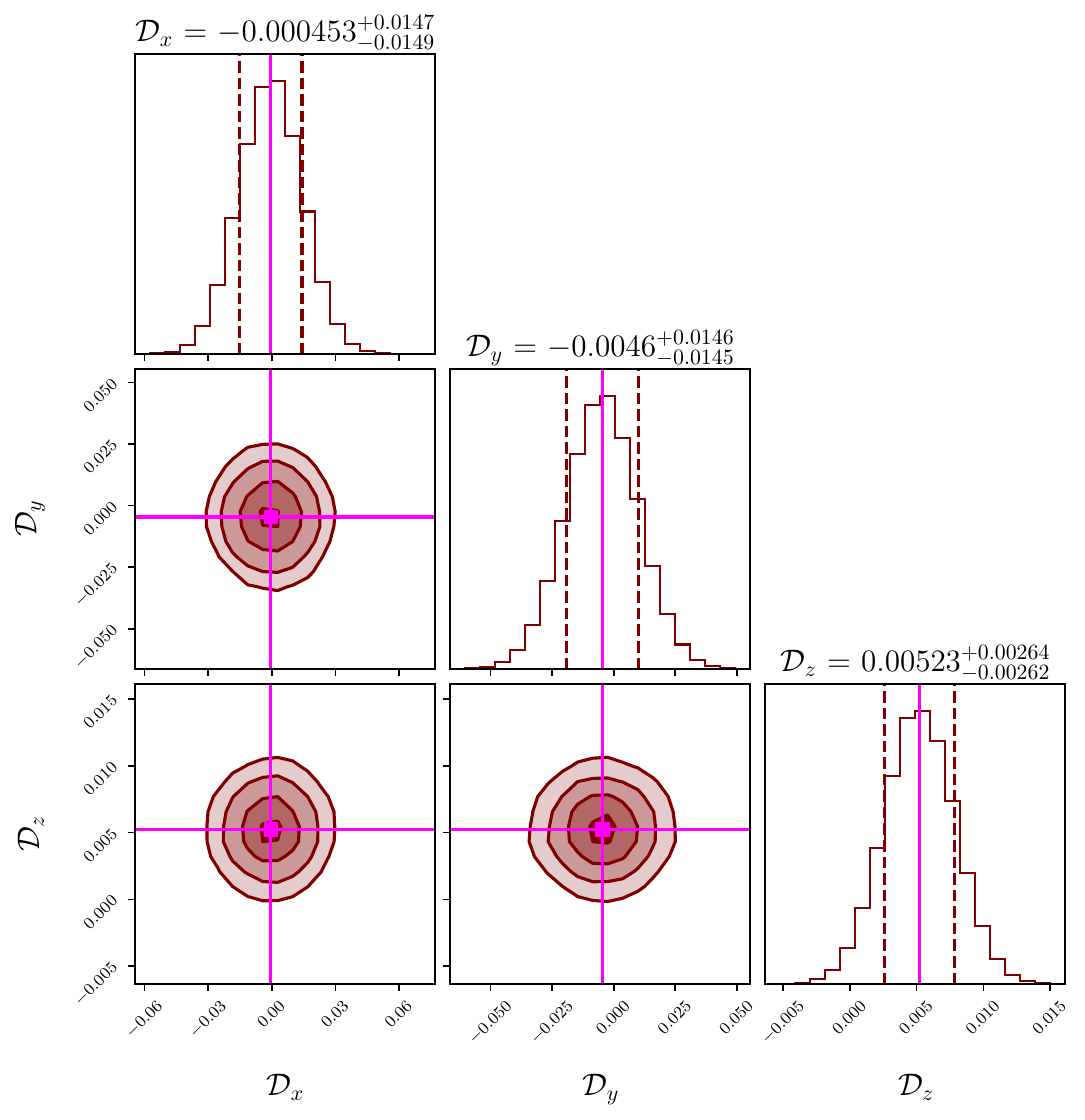}}\hfill
    \subfloat[Dipole parameters]{
    \includegraphics[width=0.49\linewidth]{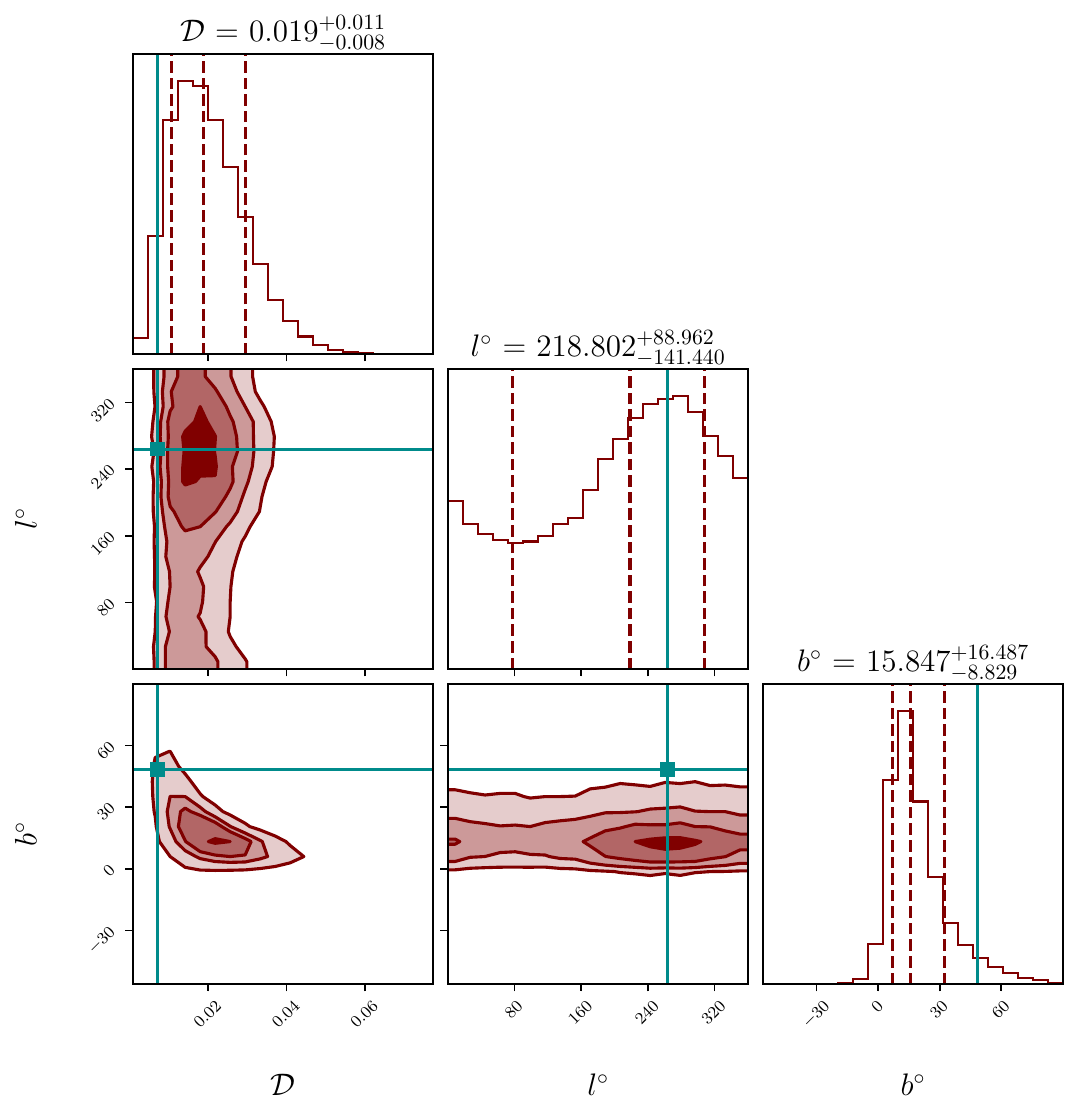}}
    \captionsetup{font=footnotesize}
    \caption{Distribution of the dipole vectors and parameters calculated using the estimator for $g_{\text{mask}}=70^{\circ}$.
    While contours in the 2D histograms enclose $12\%$, $39\%$, $68\%$ and $86\%$ of the distribution, they enclose $68\%$ of the distribution in the 1D histograms.
    The {\color{magenta} magenta} lines mark the values of the components of the true dipole vector, while {\color{darkcyan} dark cyan} lines mark the values of the true dipole parameters.
    A coordinate transformation of the distribution of the estimator's outputs from Cartesian to spherical coordinates distorts its Gaussian structure, which gives an incorrect estimate of the estimator's variance.}
    \label{fig:galcorners}
\end{figure*}

To elaborate on this, in Fig.~\ref{fig:galcorners} we visualise how the distribution
of estimator's outputs changes moving from Cartesian to spherical coordinates.
Here, we have isolated the case of $g_{\text{mask}}^\circ=70$.
In the left panel, we plot the 2D and 1D projected histograms of the Cartesian components ($\mathcal{D}_x$, $\mathcal{D}_y$ and $\mathcal{D}_z$) of the estimated dipole vectors.
Meanwhile, in the right panel, we make a similar plot of the spherical components ($\mathcal{D}$, $l^\circ$ and $b^\circ$) of the estimated dipole vectors.
The true values are shown by the coloured lines, with magenta representing the truths in Cartesian parameterization and dark cyan in spherical coordinates.
Strikingly, while the projected histograms for Cartesian components are Gaussians with means centred around the truths, the histograms for spherical components are considerably more complex in shape.
Their medians do not coincide with the truths, and the degeneracy in angle (the horizontal band above the Galactic equator in the $l^\circ$-$b^\circ$ 2D distribution) is readily apparent.
This means that one would make biased inferences about the dipole parameters.
For example, the 1D projected histogram for the amplitude shows a preference for larger amplitudes while suppressing those between 0 and the truth of 0.007. 
Thus, if the sky coverage is low, the estimator has an intrinsic bias in spherical coordinates. 
Furthermore, while the estimator's output in the Cartesian parameterisation is Gaussian, a transformation to spherical parameterisation transforms this Gaussian structure.
Since this transformation is nonlinear, the transformed distribution does not reflect the correct uncertainty in the estimator.
This means that the correct method of estimating an estimator's variance is to use the results from mock catalogues to first calculate the covariance matrix in Cartesian coordinates and then convert it to spherical coordinates through a Jacobian transformation.
If we perform this exercise with the estimator's output, we find that although the magnitude of the estimator's variance decreases for all mask types, a decrease in sky coverage still results in an increase in variance.

Given this scenario, one might wonder about whether the estimator is a good choice for calculating the dipole.
To answer this, we need to study both bias and variance in conjunction.
We can see that in the case of sufficient sky coverage, the estimator is unbiased and the variance in its results is relatively low.
This means that the estimator is a good mathematical tool for calculating the dipole.
When the sky coverage decreases, even though the estimator's results remain unbiased, its variance starts to increase significantly, which indicates that another estimator might be a better choice in this regime.

How does the estimator compare with the Bayesian approach?
Recall that the posterior distributions for the dipole directions and amplitudes are illustrated in Figures 7--8 of \citet{Oayda:2024voo}, and by the maroon contours and histograms in Figure \ref{fig:eqmask} respectively.
Looking at these plots, one could argue that the Bayesian approach suffers from a similar issue: the inferred dipole direction drifts towards the Galactic equator as the Galactic plane mask increases.
There are, however, two key differences.
First, larger mask angles are correlated with a widening of the amplitude marginal distribution and an increase in probability density for near-zero dipole amplitudes.
This means a dipole with near-zero amplitude is favoured more than one with an amplitude much larger than the truth.
The converse is true for the estimator.
The tendency to develop a long tail in the high-amplitude region likely stems from power leakage from higher multipoles into the dipole amplitude. 
This effect is exacerbated by an increase in the masked sky fraction \citep{abghari2024}.
Second, lower sky coverages (larger masks) are correlated with lower Bayes factors for the dipole versus the monopole.
Indeed, by $g_{\text{mask}}^\circ=70$, the dipole and monopole odds ratio is $\approx 1$; neither model has more explanatory power than the other.
This is an important conclusion since we would be less confident that a dipole can account for the data.
These differences imply that if sky coverage is sufficient, then both the estimator and Bayesian inference give us similar and reliable results.
But if the sky coverage is low, then due to an increase in its variance, the estimator gives us unreliable results.
Bayesian inference is often inconclusive in this regime.
Here, low Bayes factors imply that the data is insufficient to support the presence of a dipole as compared to a monopole, which hints towards the need for more data.

This discussion also highlights an additional benefit: Bayes factors allow us to compare very disparate models while conserving the basic form of the likelihood function (e.g. its pixelated functional form).
For example, the double dipole model in \citet{mittal2024} was a useful diagnostic tool for understanding the effects of extinction in the Quaia catalogue, an optical sample of quasars \citep{storeyfisher2023quaia}, whereby a region of elevated source density near the Galactic Centre could be accounted for by inadequacies in the selection function.
Of course, the \textsc{healpy} estimator is constructed purely assuming a dipole model, so asking more of it than this is not a fair comparison.
Even so, it is important to note that one would have to use a different estimator altogether to analyse more complex models.
As a second example, consider a template catalogue modulated by a dipole and additional higher-order multipoles, with a substantial sky mask applied. 
For such catalogues, if we use an extended version of the \textsc{healpy} estimator to simultaneously estimate all multipoles, then, as demonstrated by \citet{abghari2024}, the dipole amplitude sees a significant contribution from higher multipoles.
Bayesian inference avoids this leakage problem and enables robust inference of the parameters associated with all multipoles \citep{Oayda:2024voo}—as long as one includes
those multipoles in the model.

In addition, our Bayesian approach is generalisable to studies of multipole catalogues.
In cosmic dipole studies, the classic way of jointly analysing samples has been to combine the two catalogues to create a single sample \citep{colin2017}.
This is achieved by scaling the number density and flux of one catalogue to match the other.
The median spectral index of the sources, common in both catalogues, is used to scale the flux densities of one of the catalogues.
This resultant sample is used to determine the matter dipole, as has been done in radio galaxy studies such as \citet{colin2017} and \citet{darling2022}.
But, as shown in \citet{wagenveld2023}, flux scaling can introduce systematic errors, leading to spurious dipole signals.
Instead, \citet{secrest2022} calculated the dipoles in two datasets individually, showed that the results are consistent with each other, and finally calculated the joint statistical significance of their results.
However, this is not a true joint analysis, because the dipole in each catalogue is being calculated independently of the other one.
Bayesian inference, on the other hand, allows us to construct a joint likelihood function that incorporates information from all the catalogues analysed.
We can then assign different amplitudes and a common set of directions to the dipoles in each sample, computing the joint posterior distribution for these parameters and a set of Bayes factors ranking competing models.
This approach has been employed for two radio catalogues in one of our previous studies \citep{oayda2024} as well as in a couple of other studies \citep{wagenveld2023, Wagenveld:2025ewl} but in the future could be extended to others across multiple wavelengths.
That being said, in a joint analysis concern has to be given to the statistical (in)dependence between
datasets.

Although a true joint analysis of multiple catalogues has not been performed using estimators till now, such a joint analysis is possible if we use the Maximum Likelihood Estimator (MLE) by defining a joint likelihood function.
\footnote{As discussed in Section \ref{sec: framework}, \textsc{healpy}'s \texttt{fit\_dipole} is a special case of MLE, constructed by defining a Gaussian likelihood function for a single dataset.}
Accordingly, a generalized version of the MLE, which is suitable for joint analysis of multiple datasets, will have the following mathematical form:
\begin{equation}
    \phi_{\text{MLE}} = \arg \: \underset{\phi}{\max} \: \mathcal{L}_{\text{joint}}\left(\phi\right)
\end{equation}
The results of this estimator coincide with the peak of Bayesian posteriors when both of them use the same likelihood function, and the prior function in the Bayesian framework is uniform and sufficiently broad.
In this study, we have used uniform priors, which makes MLE and Bayesian inference functionally equivalent in the sense that the MLE's estimate will coincide with the posterior mode under the same likelihood.
However, Bayesian inference is still slightly advantageous over this generalized form of MLE due to the following reasons.
First, Bayesian methods allow for a more principled handling of nuisance parameters.
In Bayesian methods, we determine the credible intervals for parameters of interest by marginalising over the nuisance parameters.
This is advantageous over MLE, where nuisance parameters are accounted for through the profile likelihood method, which sometimes underestimates the uncertainties in nuisance parameters.
Second, Bayesian analysis utilises priors, which integrate extra information into the analysis, and hence disfavour illogical solutions.
This regularising effect is absent in the MLE.
Due to these reasons, Bayesian methods are better suited for joint analysis of multiple catalogues, as compared to point estimators.

Finally, Bayesian inference differs from point estimators in terms of the outputs produced after analysing a single dataset.
For a single catalogue, the estimator only gives a single dipole vector as an output, while Bayesian analysis gives a probability distribution of a range of dipole vectors which are allowed by the data.
Since we only have a handful of astronomical catalogues that can be used for dipole determination, having a probability distribution for a possible range of dipole vectors offers a much deeper insight into the data being analysed.

\section{Conclusion}\label{sec: conclusion}
In this work, we have studied the impact of our choice of mathematical tool on the inference of the cosmic dipole from the sky surveys.
We first assessed the distribution of output dipoles generated by the \textsc{healpy}'s \texttt{fit\_dipole} estimator.
We found that this estimator gives an unbiased estimate for the dipole vector regardless of noise levels and sky coverage.
However, the variance in its estimates increases with a decrease in sky coverage, which leads to wide confidence intervals and reduced reliability.
We then turned to Bayesian methods, which provide both a probability distribution for the dipole parameters and the evidence, quantifying the level of support for a dipole model.
We showed that if the sky coverage is sufficient, then the peaks of the marginalized distributions coincide with the true dipole parameters.
On the other hand, in case of low sky coverage, low Bayes factors act as a safeguard against over-interpretation by indicating that the results are inconclusive, and hint towards either formulating a different model or acquiring more data.

These findings offer fresh insights into the deployment plans for the statistical techniques used to measure the cosmic dipole.
We find that the bias and variance of the estimator must be computed in Cartesian coordinates because Cartesian dipole vectors take into account both amplitudes and directions together, which ensures that all aspects of the dipole are captured.
Although the estimator's results are quite precise, yet, for a given dataset, it is just a single output, and it does not offer any additional insights into the dataset.
This suggests the need to use an optimal estimator (having low bias and variance) formulated in an appropriate coordinate system for dipole determination.
Bayesian posterior probability distributions are not plagued by the parameterisation problem, and give the correct results without any extra processing.
Here, the result is not a single datapoint; rather, it is a probability distribution of the range of output dipoles consistent with the data.
Many times, the shape of the posterior depends on the choice of likelihood functions and the prior probability distributions, which suggests the need to develop likelihood functions that more accurately reflect the observable being modelled and more restricted priors.

We want to point out that although our results have been presented for a single number density and dipole amplitude, the qualitative features discussed above are quite general and will hold for other choices as well, while the quantitative aspects can change due to a change in catalogue structure.
Overall, our work underscores the importance of principled statistical analysis in cosmic dipole analysis and lays the groundwork for more accurate dipole estimation from upcoming sky surveys.

\bigskip
\section*{Acknowledgements}
We thank the anonymous referees for their insightful feedback that improved this paper.
VM is supported by the University of Sydney's Physics Foundation Scholarship.
OTO is supported by the University of Sydney Postgraduate Award.
This work made use of the \textsc{python} packages
\textsc{dynesty} \citep{skilling2004, skilling2006, dynesty-v2.1.3-software}, \textsc{healpy} \citep{Gorski2005,Zonca2019},
\textsc{numpy} \citep{harris2020},
\textsc{matplotlib} \citep{hunter2007},
\textsc{scipy} \citep{scipy2020} and \textsc{astropy} \citep{astropy2022}.

\section*{Data and Software Availability}
The data used in this study will be made available with a reasonable request to the authors.

\bibliographystyle{bib-style}
\bibliography{Multipoles-II}
\appendix
\section{Analytical results for $\textsc{Healpy}.\texttt{fit\_dipole}$} 
 \label{sec:appendix}
We can recast the \ref{estimator-eqn-recasted} equations 
into the matrix equation $MA=B$, where $M$ is a 4x4 matrix encoding the knowledge of sky coverage, $A$ is a 4x1  free-parameter matrix, and $B$ is a 4x1 matrix encoding the data properties.
\begin{subequations}
\begin{equation}\label{3(a)}
    \begin{bmatrix}
    \sum_{i} & \sum_{i}x_{i} & \sum_{i}y_{i} & \sum_{i}z_{i}\\
    \sum_{i}x_{i} & \sum_{i}x_{i}^{2} & \sum_{i}x_{i}y_{i} & \sum_{i}x_{i}z_{i}\\
    \sum_{i}y_{i} & \sum_{i}x_{i}y_{i} & \sum_{i}y_{i}^{2} & \sum_{i}z_{i}y_{i}\\
    \sum_{i}z_{i} & \sum_{i}x_{i}z_{i} & \sum_{i}y_{i}z_{i} & \sum_{i}z_{i}^{2}\\
    \end{bmatrix}
    \begin{bmatrix}
        A_{0}\\ A_{1}\\ A_{2}\\ A_{3}
    \end{bmatrix} = 
    \begin{bmatrix}
        \sum_{i}\mathcal{N}_i\\ \sum_{i}\mathcal{N}_ix_{i}\\ \sum_{i}\mathcal{N}_{i}y_{i}\\ \sum_{i}\mathcal{N}_iz_{i}
    \end{bmatrix}
\end{equation}
\begin{equation}\label{3(b)}
        \Rightarrow MA = B \Rightarrow A = M^{-1}B
\end{equation}
\end{subequations}
Using this matrix equation, we will infer the estimator's outputs for both noiseless and noise injected catalogues

\subsubsection{Full-sky noiseless catalogue}
In the case of a full sky noiseless catalogue, the estimator recovers the correct value of the modulating dipole.
The first step in understanding this analytically is to reduce the $M$ matrix.
Consider the $\Sigma_i$ term, which is the number of pixels $N$.
Next, consider the $\Sigma_i x_i$, $\Sigma_i y_i$ and $\Sigma_i z_i$ terms.
These are $0$ because there are equal numbers of pixels on either side of coordinate planes for a pixelated sky with a sufficient number of pixels.
Finally, the $\Sigma_i x_i y_i$, $\Sigma_i y_i z_i$ and $\Sigma_i x_i z_i$ terms are also $ 0$ by similar symmetry arguments about the coordinate planes.
A more rigorous way of proving this is by performing the analysis in spherical coordinates and summing over sine and cosine terms for a finite but significant number of small steps. 
In either case, the expressions are $\approx 0$.
These reductions convert $M$ into a diagonal matrix, and we get the following expression for $A$
\begin{equation}\label{6(c)}
    \begin{bmatrix}
        A_{0} & A_{1} & A_{2} & A_{3}
    \end{bmatrix}^{T} = 
     \begin{bmatrix}
        \frac{\sum_{i}\mathcal{N}_i}{N} &
        \frac{\sum_{i}\mathcal{N}_i x_{i}}{\sum_{i}x_{i}^{2}} &
        \frac{\sum_{i}\mathcal{N}_i y_{i}}{\sum_{i}y_{i}^{2}} &
        \frac{\sum_{i}\mathcal{N}_i z_{i}}{\sum_{i}z_{i}^{2}}
    \end{bmatrix}^{T}
\end{equation}
The next step is simplifying the numerator terms on the right-hand side of the above equation. 
Consider the number density of the $i^{th}$ pixel pointing in the direction $(x_i,y_i,z_i)$.
Suppose we assume that the catalogue has a dipole of amplitude $\mathcal{D}$ pointing towards the unit vector $(x_0,y_0,z_0)$. 
Then, the number density of this pixel is $\mathcal{N}_{i} = \bar{\mathcal{N}}(1+\mathcal{D} \cos \theta_i)$, where $\theta_i$ gives the angular offset between the dipole and pixel directions.
Accordingly,
\begin{equation}
\mathcal{N}_{i} = \bar{\mathcal{N}}(1+\mathcal{D} \cos \theta_i) = \bar{\mathcal{N}} (1+\mathcal{D} x_{0}x_{i}+\mathcal{D} y_{0}y_{i}+\mathcal{D} z_{0}z_{i})
\end{equation}
Summation over all the pixels gives
\begin{subequations}
\begin{align}
    \sum_{i}\mathcal{N}_i
        &= \bar{\mathcal{N}} \left(\sum_{i}1+\mathcal{D} x_{0}\sum_{i}x_{i}+\mathcal{D} y_{0}\sum_{i}y_{i}+\mathcal{D} z_{0}\sum_{i}z_{i}\right) \\
        &= \bar{\mathcal{N}} N \nonumber \\
    \intertext{since again the terms $\sum x_i$, etc. are $\approx 0$. Meanwhile, we have}
    \sum_{i}\mathcal{N}_i x_{i}
        &= \bar{\mathcal{N}} \left[\sum_{i}x_{i}+ \allowbreak \mathcal{D} \left(x_{0}\sum_{i}x_{i}^{2}+ y_{0}\sum_{i}x_{i}y_{i}+z_{0}\sum_{i}x_{i}z_{i}\right) \right] \\
        &= \bar{\mathcal{N}}\mathcal{D} x_{0}\sum_{i}x_{i}^{2}
            \implies \frac{\sum_{i}\mathcal{N}_{i}x_{i}}{\sum_{i}x_{i}^{2}}
            = \bar{\mathcal{N}}\mathcal{D} x_{0}
    \end{align}
\end{subequations}
with similar results for $\sum_{i}\mathcal{N}_i z_{i}$ and $\sum_{i}\mathcal{N}_i z_{i}$. 
The net dipole vector is therefore given by $\mathcal{G} = \mathcal{D}(x_0,y_0,z_0)$, which shows that the estimator is unbiased in the noiseless case.
We have verified this statement through simulations as well.
We have found that the dipole vector inferred from all-sky noiseless mock catalogues agrees with the true dipole vector used for generating the catalogue, regardless of the dipole parameters used to generate the catalogue.

\subsubsection{Masked noiseless catalogue}
Masking subtracts a portion of the sky from the full-sky map, affecting both the $M$ and $B$ matrices. This can inject bias into the sample.
For our purpose, masking removes a set of pixels $\{k\}$ from the summation over all the pixels, i.e. $\sum_{i\neq k} = \sum_{i} - \sum_{k}$.
This changes the $M$ matrix to
\begin{subequations}
   \begin{equation}
    M = \begin{bmatrix}
    \sum_{i}-\sum_{k} & -\sum_{k}x_{k} & -\sum_{k}y_{k} & -\sum_{k}z_{k}\\
    -\sum_{k}x_{k} & \sum_{i}x_{i}^{2}-\sum_{k}x_{k}^{2} & -\sum_{k}x_{k}y_{k} & -\sum_{k}x_{k}z_{k}\\
    -\sum_{k}y_{k} & -\sum_{k}x_{k}y_{k} & \sum_{i}y_{i}^{2}-\sum_{k}y_{k}^{2} & -\sum_{k}z_{k}y_{k}\\
    -\sum_{k}z_{k} & -\sum_{k}x_{k}z_{k} & -\sum_{k}y_{k}z_{k} & \sum_{i}z_{i}^{2}-\sum_{k}z_{k}^{2}\\
    \end{bmatrix}
    \end{equation}
    and the $B$ matrix to
    \begin{equation}
    B = \begin{bmatrix}
        \sum_{i}\mathcal{N}_{i}-\sum_{k}\mathcal{N}_{k}\\ 
        \sum_{i}\mathcal{N}_{i}x_{i}-\sum_{k}\mathcal{N}_{k}x_{k}\\ 
        \sum_{i}\mathcal{N}_{i}y_{i}-\sum_{k}\mathcal{N}_{k}y_{k}\\ 
        \sum_{i}\mathcal{N}_{i}z_{i}-\sum_{k}\mathcal{N}_{k}z_{k}
    \end{bmatrix}.
    \end{equation} 
\end{subequations}
Subsequent calculations will show that in the case of arbitrarily masked skies, the magnitude of any component of the dipole vector will depend on all the cartesian components of the masked pixels.
As an example, $\mathcal{D}_x$ will depend on the set of all $x_k$, $y_k$ and $z_k$.

Some simplification, however, is achievable if masking removes equal regions above and below the Galactic plane.
Then, due to the symmetry of masked pixels around the coordinate planes, the off-diagonal elements of the $M$ matrix are $\approx 0$:
\begin{equation}
    A = \begin{bmatrix}\label{6(a)}
    N-N' & 0 & 0 & 0\\
    0 & \sum_{i}x_{i}^{2}-\sum_{k}x_{k}^{2} & 0 & 0\\
    0 & 0 & \sum_{i}y_{i}^{2}-\sum_{k}y_{k}^{2} & 0\\
    0 & 0 & 0 & \sum_{i}z_{i}^{2}-\sum_{k}z_{k}^{2}\\
    \end{bmatrix}
\end{equation}
The $x$ component of the inferred dipole becomes
\begin{equation}\label{eq:galmaskxtermestimator}
    \Rightarrow \frac{A_1}{A_0} = \mathcal{D} x_{0}
    \frac{\left(1-\frac{\sum_{k}\mathcal{N}_{k}x_{k}}{\bar{\mathcal{N}}\mathcal{D} x_{0}\sum_{i}x_{i}^{2}}\right) \left(1-\frac{N'}{N}\right)} {\left(1-\frac{\sum_{k}x_{k}^{2}}{\sum_{i}x_{i}^{2}}\right)\left(1-\frac{\sum_{k}\mathcal{N}_{k}}{\bar{\mathcal{N}}N}\right)}
\end{equation}
With similar equations for the $y$ and $z$ components. 
Finally, the modified dipole parameters $(\mathcal{D}_{\text{masked}}, \theta_{\text{masked}}, \phi_{\text{masked}})$ are:
\begin{subequations}\label{eq:galmaskestimator}
    \begin{equation}
        \mathcal{D}_{\text{masked}} = \sqrt{\left(\frac{A_1}{A_0}\right)^2 + \left(\frac{A_2}{A_0}\right)^2 + \left(\frac{A_3}{A_0}\right)^2}
    \end{equation}
    \begin{equation}
    \tan(\phi_{\text{masked}}) = \frac{y_0}{x_0}\frac{\left(1-\frac{\sum_{k}\mathcal{N}_{k}y_{k}}{\bar{\mathcal{N}}\mathcal{D} y_{0}}\right)}{\left(1-\frac{\sum_{k}\mathcal{N}_{k}x_{k}}{\bar{\mathcal{N}}\mathcal{D} x_{0}}\right)} 
    = \frac{\left(\bar{\mathcal{N}}\mathcal{D} y_{0}\sum_{i}y_{i}^{2}-\sum_{k}\mathcal{N}_{k}y_{k}\right)}{\left(\bar{\mathcal{N}}\mathcal{D} x_{0}\sum_{i}x_{i}^{2}-\sum_{k}\mathcal{N}_{k}x_{k}\right)}
\end{equation}
\begin{equation}
    \cos(\theta_{\text{masked}}) = \frac{\mathcal{D} z_{0}}{\mathcal{D}_{\text{masked}}}
      \frac{\left(1-\frac{\sum_{k}\mathcal{N}_{k}z_{k}}{\bar{\mathcal{N}}\mathcal{D} z_{0}\sum_{i}z_{i}^{2}}\right) \left(1-\frac{N'}{N}\right)} {\left(1-\frac{\sum_{k}z_{k}^{2}}{\sum_{i}z_{i}^{2}}\right)\left(1-\frac{\sum_{k}\mathcal{N}_{k}}{\bar{\mathcal{N}}N}\right)}
\end{equation}
\end{subequations}
Further analytical simplification is not possible in this case.
We deduce that in the presence of a Galactic plane mask, the magnitude of any component of the dipole vector only depends on the corresponding cartesian coordinate of the masked pixels i.e. $\mathcal{D}_x$ depends only on $\{x_k\}$, $\mathcal{D}_y$ depends only on $\{y_k\}$ and $\mathcal{D}_z$ depends only on $\{z_k\}$.
Although making an analytical deduction of bias in this case is not feasible, some progress is achievable if one relies on simulations.

Using simulations, we have again deduced that the calculated dipoles are equal to the dipole used for modulating the template catalogue, regardless of the dipole parameters used to generate the catalogue, as well as the shape and the size of the mask.

\subsubsection{Catalogues with noise}
\begin{figure}
    \centering
    \includegraphics[width=\linewidth]{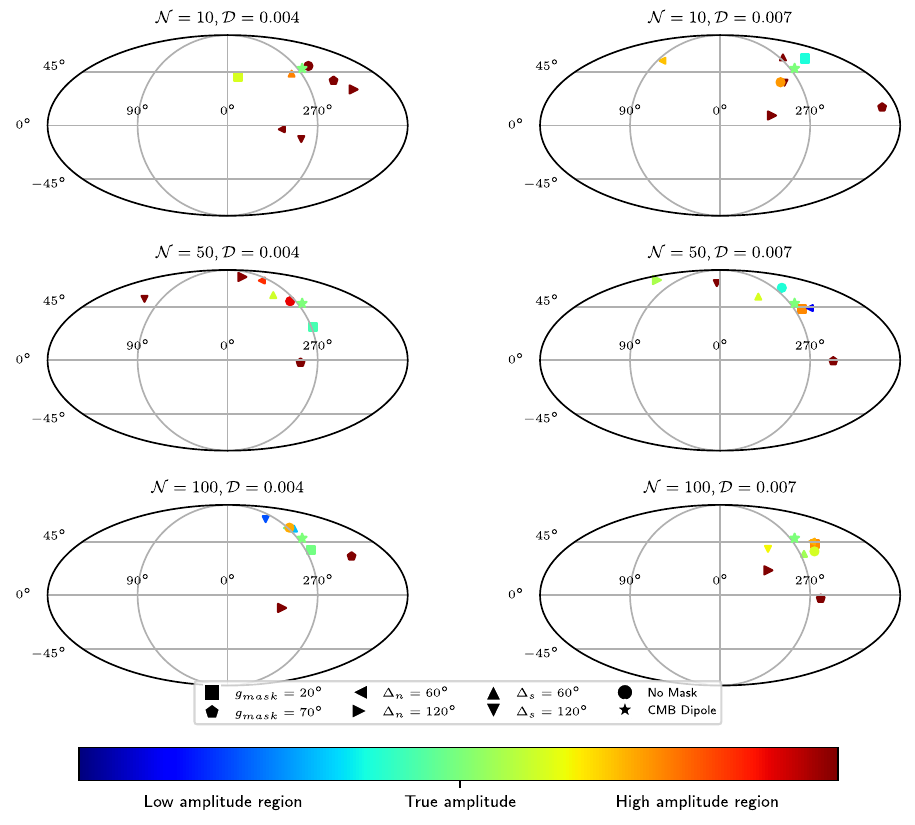}
    \captionsetup{font=footnotesize}
    \caption{Estimator's output vectors (converted to dipole parameters) for noise-injected mock catalogues. Each vector has been calculated using a single catalogue.}
    \label{fig:nosiy}
\end{figure}

The presence of noise in number density drastically changes the situation, as it modifies the number density of the catalogue $\mathcal{N}_i + \delta \mathcal{ N}_i$.
This means that the $B$ matrix gains an additional component $\delta B$
\begin{equation}\label{8}
    B +\delta B = 
    \begin{bmatrix}
        \sum_{i}\mathcal{N}_i\\ \sum_{i}\mathcal{N}_i x_{i}\\ \sum_{i}\mathcal{N}_i y_{i}\\ \sum_{i}\mathcal{N}_i z_{i}
    \end{bmatrix} +
    \begin{bmatrix}
        \sum_{i}\delta \mathcal{ N}_i\\ \sum_{i}\delta \mathcal{ N}_i x_{i}\\ \sum_{i}\delta \mathcal{ N}_i y_{i}\\ \sum_{i} \delta \mathcal{ N}_i z_{i}
    \end{bmatrix}
\end{equation}
This adds a component $\delta A$ and, hence, a fluctuation vector $\delta \mathcal{G}$ to the dipole vector $\mathcal{G}$
\begin{subequations}
    \begin{equation}
        \delta A = 
        \begin{bmatrix}
        \frac{\sum_{i}\delta \mathcal{ N}_i}{N} &
        \frac{\sum_{i}\delta \mathcal{ N}_i x_{i}}{\sum_{i}x_{i}^{2}} &
        \frac{\sum_{i}\delta \mathcal{ N}_i y_{i}}{\sum_{i}y_{i}^{2}} &
        \frac{\sum_{i}\delta \mathcal{ N}_i z_{i}}{\sum_{i}z_{i}^{2}}
    \end{bmatrix}^T
    \end{equation}
    \begin{equation}
        \delta \mathcal{G} = 
        \begin{bmatrix}
        \frac{N\sum_{i}\delta \mathcal{ N}_i x_{i}}{\sum_{i}x_{i}^{2} \sum_{i}(\mathcal{ N}_i + \delta \mathcal{ N}_i)} &
        \frac{N\sum_{i}\delta \mathcal{ N}_i y_{i}}{\sum_{i}y_{i}^{2} \sum_{i}(\mathcal{ N}_i + \delta \mathcal{ N}_i)} &
        \frac{N\sum_{i}\delta \mathcal{ N}_i z_{i}}{\sum_{i}z_{i}^{2} \sum_{i}(\mathcal{ N}_i + \delta \mathcal{ N}_i)}
    \end{bmatrix}^T
    \end{equation}
\end{subequations}
For a full-sky map, since $\sum_{i}x_{i}^{2}=\sum_{i}y_{i}^{2}=\sum_{i}z_{i}^{2}$ and $N$ is fixed, the fluctuation vector is reduced to
\begin{equation}
        \delta \mathcal{G} = \frac{N}{\sum_{i}x_{i}^{2} \sum_{i}(\mathcal{ N}_i + \delta \mathcal{ N}_i)}
        \begin{bmatrix}
        \sum_{i}\delta \mathcal{ N}_i x_{i} &
        \sum_{i}\delta \mathcal{ N}_i y_{i} &
        \sum_{i}\delta \mathcal{ N}_i z_{i}
    \end{bmatrix}^T
\end{equation}\label{eq:delG_fullsky}
A similar analysis for Galactic plane masked leads to the following expression for $ \delta \mathcal{G}$ 
\begin{equation}
        \delta \mathcal{G} = \frac{N-N'}{\sum_{i\neq k}(\mathcal{ N}_i + \delta \mathcal{ N}_i)}
        \begin{bmatrix}
        \frac{\sum_{i\neq k}\delta \mathcal{ N}_i x_{i}}{ \sum_{i\neq k}x^2_i} &
        \frac{\sum_{i\neq k}\delta \mathcal{ N}_i y_{i}}{ \sum_{i\neq k}y^2_i} &
        \frac{\sum_{i\neq k}\delta \mathcal{ N}_i z_{i}}{ \sum_{i\neq k}z^2_i}
    \end{bmatrix}^T
\end{equation}\label{eq:delG_gplmasky}
We can use these equations to determine the dipole vector and its parameters.
Since an analytical approach is not feasible here, we look at the results from simulations.
Fig \ref{fig:nosiy} shows the dipole parameters $(\mathcal{D}, l^{\circ}, b^{\circ})$ calculated from the estimator's output vectors for both full and masked skies at different number densities, again imprinted with dipoles of varying amplitudes pointing towards the CMB dipole.
Each output vector in the figure has been calculated over a single catalogue.
Additionally, noise has been injected in each dataset using the method discussed in Section \ref{sec: methodology}.
We see that noise injection shifts the dipole's amplitude and direction away from its true value.
The shift depends not only on the number density of the catalogue but also on the shape and size of the mask, as seen by the difference in inferred vectors and parameters for different number densities and mask types.

\section{Comparing the estimator results}
\label{sec:estimator-bias-comparison}
\begin{figure*}[!ht]
    \centering
    \subfloat[$\Delta_n=60^{\circ}$]{
    \includegraphics[width=0.49\linewidth]{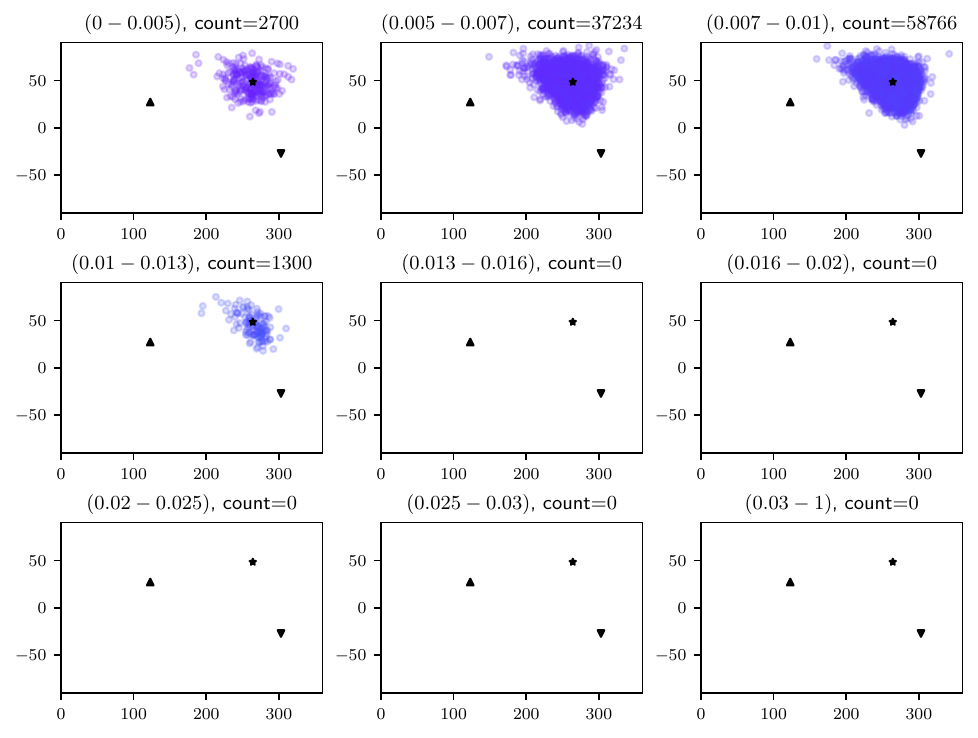}}\hfill
    \subfloat[$\Delta_n=120^{\circ}$]{
    \includegraphics[width=0.49\linewidth]{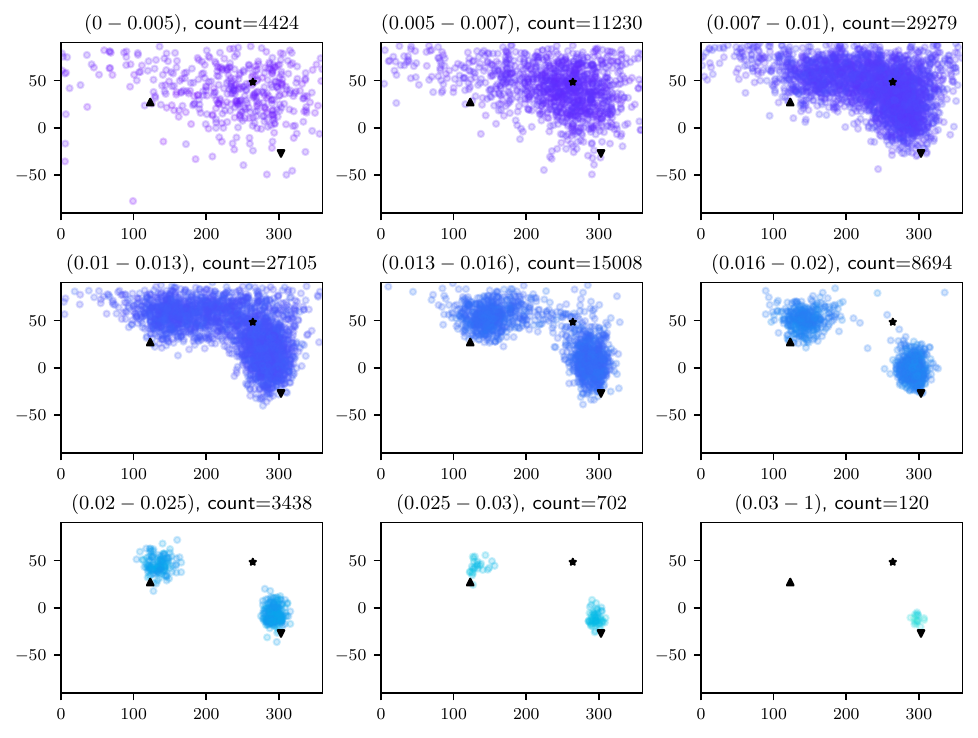}}\\
    \subfloat[$g_{mask}=30^{\circ}$]{
    \includegraphics[width=0.49\linewidth]{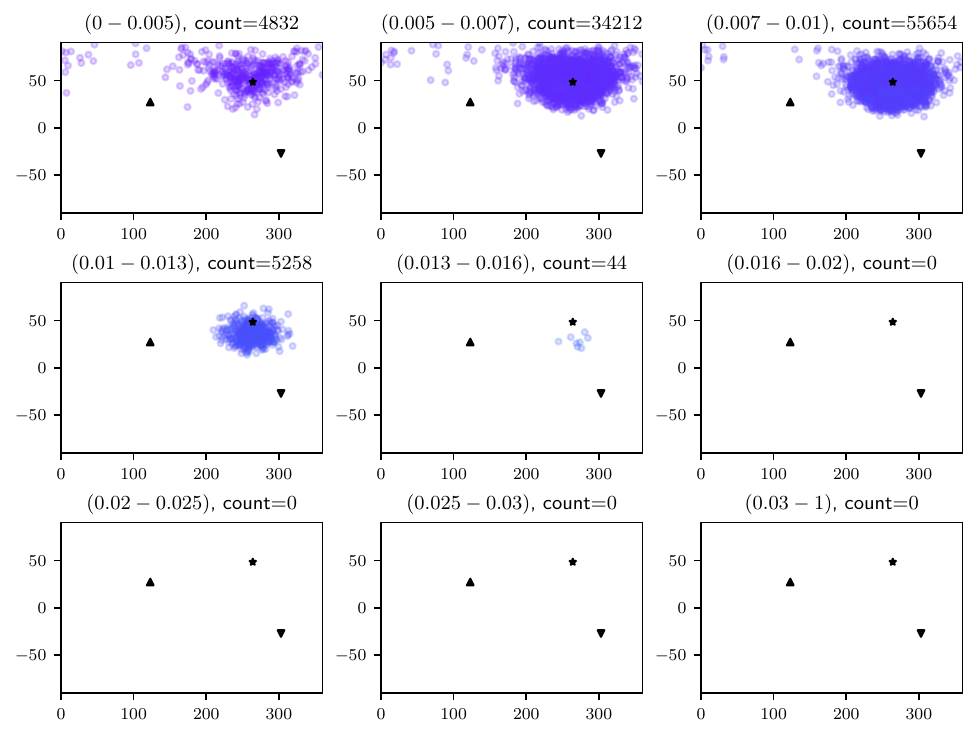}}\hfill
    \subfloat[$g_{mask}=70^{\circ}$]{
    \includegraphics[width=0.49\linewidth]{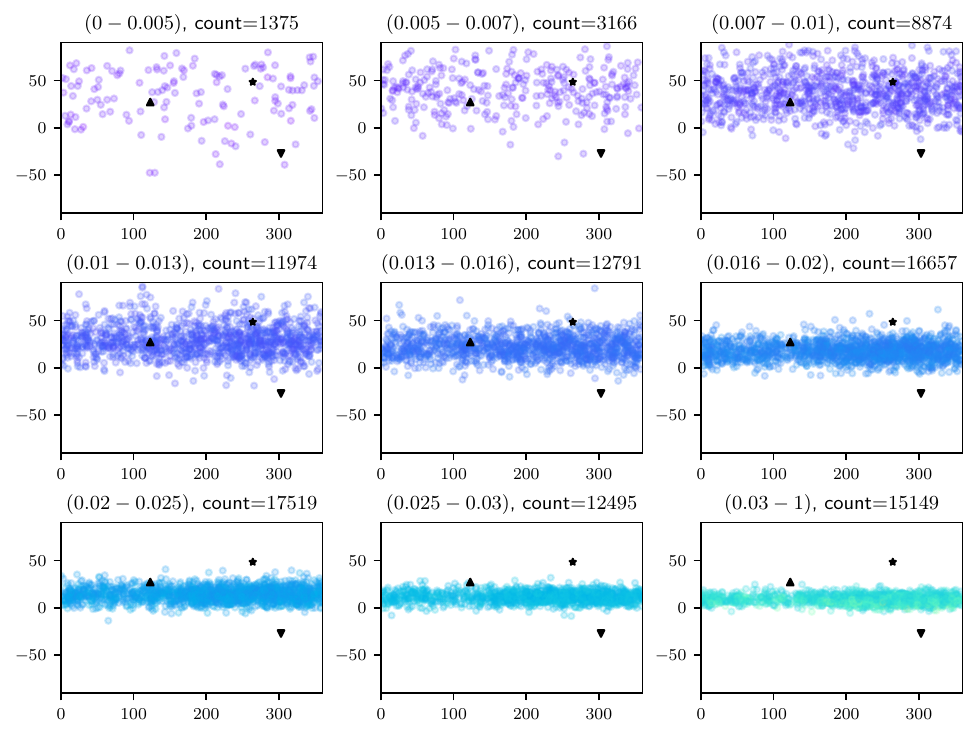}}\\
    \captionsetup{font=footnotesize}
    \caption{Distribution of calculated dipole directions within different amplitude bins for a number of different masks.
    For all the plots, $l^{\circ}$ is along the $x$-axis,while $b^{\circ}$ is along the $y$-axis. 
    $\filledstar$ represents the CMB dipole direction, $\blacktriangle$ the north equatorial pole, and $\blacktriangledown$ the south equatorial poles.
    The title of each subplot uses the following convention:
        (lower amplitude limit -- higher amplitude limit),
        count = number of dipole vectors within the bin.
    \textit{Top row:} $60^{\circ}$ and $120^{\circ}$ cap masks around the north equatorial poles.
    \textit{Bottom row:} $30^{\circ}$ and $70^{\circ}$ masks around the Galactic plane.
    }
    \label{fig:ampslice_freq}
\end{figure*}
The difference in the estimator's results for different averaging methods poses the question as to which method should be used for quantifying the estimator's bias and by extension, used to determine the dipole from the dataset.
To answer this question, we need to understand the cause behind the discrepancy between the results calculated using either method.
We explore this issue in the following paragraphs.

We first look at the results for the equatorial cap masks.
To understand things better, we restrict our discussion to the north equatorial cap masks with radii $60^{\circ}$ and $120^{\circ}$.
We select these masks because, for $\Delta_n = 60^{\circ}$ mask, the output parameters for both averaging methods coincide with the true dipole parameters. 
In contrast, for the $\Delta_n = 120^{\circ}$ mask, only the dipole parameters calculated from Method 1 (vector averages) align with the true dipole parameters.
In the top row of Fig \ref{fig:ampslice_freq}, we slice the estimator's outputs into amplitude bins and plot the direction parameters $(l^{\circ},b^{\circ})$ of all the data points in each bin.
We see that for a small $\Delta_n$ (high sky coverage), only $~ 1\%$ of the dipoles have high amplitude $(\mathcal{D} > 0.01)$.
The dipole directions are uniformly scattered around the CMB dipole direction regardless of their amplitude values (note that the large spread in the distribution in the upper part of any plot is because we are projecting a sphere on a flat surface).
For a large $\Delta_n$ (low sky coverage), more than half of the distribution is present in high amplitude bins $(\mathcal{D} > 0.01)$, and the direction parameters are concentrated around the equatorial poles in roughly equal numbers.
The remaining distribution is present in the low amplitude bins.
These plots highlight a key relation between the amplitudes and directions of the recovered dipoles. 
If the sky coverage is insufficient, a dipole with high amplitude is more likely to be near one of the equatorial poles.

Using these features, we can deduce why the average over the parameters of output dipoles for a large equatorial cap mask does not coincide with the true dipole parameters.
In the presence of a large mask, most of the vectors in the dataset have high amplitude, and consequently, we see a shift in average dipole amplitude towards higher values.
We also see a clustering of direction parameters around the equatorial poles.
Hence, the average direction shifts away from the CMB dipole direction.
In other words, the shift arises because we take disjoint averages of the amplitudes and directions.
This disjoint averaging undermines the above-mentioned correlation between the amplitudes and directions for the recovered dipole.
On the other hand, any component of an output dipole vector is a function of amplitudes and directions.
Consequently, the average vector contains information about the correlation between the parameters.
All the high amplitude vectors are concentrated around the equatorial poles in roughly equal numbers. 
They cancel each other while taking the sum, reducing the amplitude and shift of the average dipole vector.

In the case of high sky coverage, nearly all of the distribution is present in the low amplitude bins $(<0.01)$ and is concentrated around the CMB dipole direction.
Furthermore, $\approx 59\%$ of the distribution has amplitude $>0.007$, while the remaining $\approx 41\%$ has amplitude $<0.007$.
Hence, disjoint averages of dipole parameters are near the true values.
It also means that the number of dipole vectors with amplitudes less than the
true value ($\mathcal{D} = 0007$) is approximately equal to the number of vectors
with a greater amplitude.
Furthermore, they are uniformly scattered around the CMB dipole direction.
Consequently, along any given direction, the high-amplitude dipole vectors will cancel the effect of low-amplitude dipoles.
Additionally, a uniform scatter of vector directions averages out to the CMB dipole direction, which brings the average vector nearer to the true dipole vector.

We now look at the results for Galactic plane masks.
Similar to the equatorial cap masks, we restrict our discussion to the results for $g_{\text{mask}} = 30^{\circ}$ and $70^{\circ}$ masks.
These choices have been motivated by the agreement between the outputs of both averaging methods and true dipole parameters for a $30^{\circ}$ mask and their disagreement for the $70^{\circ}$ mask.
The bottom rows of Fig \ref{fig:ampslice_freq} show the corresponding sliced dipole distributions.
We see that for the high sky coverage (low $g_{\text{mask}}$), nearly all of the dipoles have low amplitudes $(<0.01)$ and all the dipole directions are uniformly scattered around the CMB dipole direction.
Meanwhile, for the low sky coverage (high $g_{\text{mask}}$), $\approx 87\%$ of the dipole distribution is present in the high amplitude bins, and the corresponding directions are scattered in a uniform strip above the Galactic equator.
We infer another relation between the output dipole parameters for a large mask: high amplitude dipoles have a lower angular offset with the Galactic equator.

\begin{table}
    \centering
    \setlength{\tabcolsep}{20pt}
    \renewcommand{\arraystretch}{1.2}
    \begin{tabular}{c c c}
        \toprule
         Amplitude Bin & Data Points & $\bar{\mathcal{D}_z}$ \\
         \midrule
        $0.000-0.005$ & $1375$ & $0.00218$\\
        \midrule
        $0.005-0.007$ & $3166$ & $0.00369$\\
        \midrule
        $0.007-0.010$ & $8874$ & $0.00478$\\
        \midrule
        $0.010-0.013$ & $11974$ & $0.00529$\\
        \midrule
        $0.013-0.016$ & $12791$ & $0.0054$\\
        \midrule
        $0.016-0.020$ & $16657$ & $0.00537$\\
        \midrule
        $0.020-0.025$ & $17519$ & $0.00542$\\
        \midrule
        $0.025-0.030$ & $12495$ & $0.00538$\\
        \midrule
        $0.030-1.000$ & $15149$ & $0.00541$\\
        \midrule
        $0-1$ & $100000$ & $0.00523$\\
        \bottomrule
    \end{tabular}
    \captionsetup{font=footnotesize}
    \caption{The average value of $\mathcal{D}_z$ for ($\bar{\mathcal{D}_z}$) different amplitude bins for $g_{\text{mask}}=70^{\circ}$.
    The mock catalogues were constructed using $\mathcal{D}=0.007$ and $b^{\circ}=48\dotdeg253$.
    Consequently the true dipole had $\mathcal{D}_z = 0.00522$}
    \label{tab:g70binz}
\end{table}
Now, one can ask the question: why, for the $70^{\circ}$ mask, do we see a uniform strip-like distribution of high-amplitude dipoles above the Galactic equator and yet recover the true parameters by averaging over the vectors (Method 1)?
We will explain the reason behind this using the $z$-component of the dipole vector: $\mathcal{D}_z = \mathcal{D} \sin b$.
This expression tells us that a low $b^{\circ}$ (and hence a low $\sin b$) balances the high amplitude, which brings the value of $\mathcal{D}_z$ down to a lower value.
So, the average $\mathcal{D}_z$ for vectors in a high amplitude bin should approach the true $\mathcal{D}_z$, which is a small number.
In Table \ref{tab:g70binz}, we list the average $\mathcal{D}_z$ for vectors present in the amplitude bins used in Fig \ref{fig:ampslice_freq}.
We see that the average $\mathcal{D}_z$ for the high amplitude bins approaches the true value (the true dipole parameters $\mathcal{D}=0.007$ and $b^{\circ}=48\dotdeg253$ give $\mathcal{D}_z=0.00522$).
Since we have an excess of higher amplitude terms, the net $\bar{\mathcal{D}_z}$ also approaches the true value.
Similar reasoning holds for $x$ and $y$ components as well.
This balancing of terms in vector averages is absent while taking disjoint averages of output amplitude and directions, so the average amplitude shifts towards higher values, and the average direction shifts towards the Galactic equator.

In the case of high sky coverage, we again have a uniform clustering of dipoles around the CMB dipole direction and an equal number of dipoles on either side of the true amplitude.
Hence, as discussed in the case of the equatorial cap mask, we will see the equality of disjoint averages of dipole parameters, parameters calculated from vector averages and the true values.

The foregoing suggests that in the case of low sky coverage, a disjoint average over either amplitude or directions undermines the correlation between them.
Averaging over the amplitude gives us higher amplitude values, while averaging over the directions gives us sizeable angular offsets from the CMB dipole direction.
The correct approach, therefore, is to average over the output vectors in Cartesian space (sum them)
and use the resultant vector for bias analysis.
We thus find that the estimator gives unbiased estimates of the dipole vectors,
but is biased where one disjointly averages
over the output parameters in $\mathcal{D}$, ($l^\circ$, $b^\circ)$ space.
\label{lastpage}

\end{document}